\documentclass[preprint]{emulateapj}
\usepackage{natbib}
\usepackage{graphicx}

\begin{document} 

\title{EX Lupi from Quiescence to Outburst: Exploring the LTE Approach in Modelling Blended H$_{2}$O and OH Mid-Infrared Emission} 

\author{A. Banzatti\altaffilmark{1}, M. R. Meyer\altaffilmark{1}, S. Bruderer\altaffilmark{1,2}, V. Geers\altaffilmark{1}, I. Pascucci\altaffilmark{3}, F. Lahuis\altaffilmark{4}, A. Juh\'asz\altaffilmark{5,6}, T. Henning\altaffilmark{5}, P. \'Abrah\'am\altaffilmark{7}} 
\altaffiltext{1}{ETH Z\"urich, Institut f\"ur Astronomie, Wolfgang-Pauli-Strasse 27, CH-8093 Z\"urich, Switzerland} 
\altaffiltext{2}{Max-Planck-Institut f\"ur Extraterrestrische Physik, Giessenbachstr. 1, D-85748 Garching bei M\"unchen, Germany} 
\email{banzatti@astro.phys.ethz.ch}
\altaffiltext{3}{Space Telescope Science Institute, 3700 San Martin Drive, Baltimore, MD 21218, USA} 
\altaffiltext{4}{SRON Netherlands Institute for Space Research, P.O. Box 800, NL 9700 AV Groningen, The Netherlands} 
\altaffiltext{5}{Max-Planck-Institut f\"ur Astronomie, K\"onigstuhl 17, D-69117 Heidelberg, Germany} 
\altaffiltext{6}{Leiden Observatory, Leiden University, P.O. Box 9513, NL-2300 RA Leiden, The Netherlands} 
\altaffiltext{7}{Konkoly Observatory, Konkoly Thege Mikl\'os 15-17, H-1121 Budapest, Hungary}

\begin{abstract} 
We present a comparison of archival \textit{Spitzer} spectra of the strongly variable T Tauri EX Lupi, observed before and during its 2008 outburst. We analyze the mid-infrared emission from gas-phase molecules thought to originate in a circumstellar disk. In quiescence the emission shows a forest of H$_{2}$O lines, highly excited OH lines, and the Q branches of the organics C$_{2}$H$_{2}$, HCN, and CO$_{2}$, similar to the emission observed toward several T Tauri systems. The outburst emission shows instead remarkable changes: H$_{2}$O and OH line fluxes increase, new OH, H$_{2}$, and H\textsc{i} transitions are detected, and organics are no longer seen. We adopt a simple model of a single-temperature slab of gas in local thermal equilibrium, a common approach for molecular analyses of \textit{Spitzer} spectra, and derive the excitation temperature, column density, and emitting area of H$_{2}$O and OH. We show how model results strongly depend on the selection of emission lines fitted, and that spectrally-resolved observations are essential for a correct interpretation of the molecular emission from disks, particularly in the case of water. Using H$_{2}$O lines that can be approximated as thermalized to a single temperature, our results are consistent with a column density decrease in outburst while the emitting area of warm gas increases. A rotation diagram analysis suggests that the OH emission can be explained with two temperature components, which remarkably increase in column density in outburst. The relative change of H$_{2}$O and OH emission suggests a key role for UV radiation in the disk surface chemistry. 
\end{abstract} 


\section{INTRODUCTION} \label{sec:intro}
EX Lupi is a young star+disk system whose photometric variability was discovered in 1944 \citep{McL}. Since then the source has been monitored and showed repetitive  eruptive phenomena related to significant changes in the amount of accreting material onto the star \citep{leh95,herb07,attila}. Spectroscopic observations during the quiescent phases between the recurring outbursts suggest similarity with classical T Tauri stars of M0 type \citep{herb01,sip}. The discovery of similar variable sources confirmed EX Lupi as the prototype of a class of young strongly active T Tauri stars, called ``EXors" \citep{herb89,herb08}. This class of objects has been proposed as an intermediate stage between FUors and classical T Tauri stars \citep{herb77,teo99}, exhibiting repetitive outbursts that are both weaker and of shorter duration than FUors \citep{herb07,herb08}. The 2008 outburst of EX Lupi, that we study in this work, is just the most recent of the series of eruptive events recorded during the half-century-long monitoring of the source, but more interestingly it is the most extreme ever. It showed an increase of $\sim$5 mag in visual light over seven months (January--September 2008), compared to the typical increase of $\sim$1--2 mag observed in its previous characteristic outbursts \citep{asp} as well as in other T Tauri stars \citep[e.g.][]{herb}. A comparable optical brightening of the source was previously observed only in 1955, but spectroscopic observations were not performed \citep{herb77}. The 2008 event, instead, was monitored at several wavelengths over its entire duration, which has provided a deeper insight into the complex behaviour of EX Lupi and the still not well understood mechanism producing its strong outbursts \citep{abra09,asp,grosso,goto,attila,kosp11}.

\begin{deluxetable*}{l l l l l c c}
\tabletypesize{\small}
\tablewidth{500pt}
\tablecaption{\label{tab:arch}Summary of EX Lupi Data from the \textit{Spitzer} Archive (IRS Staring Mode).}
\tablehead{\colhead{Obs. ID} & \colhead{Date} & \colhead{Phase} & \colhead{Investigator (identifier)} & \colhead{Module} & \colhead{Ramp} & \colhead{t$_{int}$ ($s$) }}
\startdata
  172 & 2004-08-31 & quiescence & Evans (E04) & SH & 30 $\times$ 1 & 60\\ 
   &     & &  & LH & 60 $\times$ 1 & 120\\ 
  &     & & & SL2, SL1, LL1 & 14 $\times$ 1 & 24 \\ 
\hline  3716 & 2005-03-18  & quiescence & Stringfellow (S05) & SH & 120 $\times$ 2 & 480\\
   &     & &  & LH & 60 $\times$ 2 & 240\\ 
   &     & & & SL2, SL1 & 14 $\times$ 4 & 112 \\ 
\hline 477 & 2008-04-21  & outburst & Abraham (A08) & SH, LH, SL2, SL1 & 6 $\times$ 4 & 48\\
   &   &  &  & SH, LH (backgr.) & 6 $\times$ 4 & 48\\
\hline  50641 & 2008-05-02  & outburst & Carr (C08) & SH & 30 $\times$ 4 & 240\\
   &     & &  & LH & 60 $\times$ 4 & 480\\ 
   &     & &  & SH (backgr.) & 30 $\times$ 2 & 120\\ 
   &     & &  & LH (backgr.) & 60 $\times$ 2 & 240\\ 
\hline  524 & 2008-10-10  & quiescence & Abraham (A09) & SL2, SL1, LL2, LL1 & 6 $\times$ 4 & 48\\ 
   & 2009-04-07  &  &  & SL2, SL1, LL2, LL1 & 6 $\times$ 4 & 48  
\enddata
\end{deluxetable*}

In the last few years the \textit{Spitzer Space Telescope} \citep{spitzer} has enabled the study of the molecular content in circumstellar disks through their mid-infrared (MIR) emission, unveiling a rich chemistry in the innermost regions of T Tauri disks \citep{sal08,sal11,cn08,cn11,pasc09,pont10}. So far, such emission has been found to be common in the T Tauri systems observed \citep{pont10} and has been studied by means of simple single-slab models that assume the emitting gas to be in local thermal equilibrium (LTE). A certain diversity in the molecular emission observed toward T Tauri systems is apparent \citep{pasc09,pont10,cn11}, and disks in more evolved systems as well as around Herbig AeBe stars seem to be depleted in warm molecules compared to T Tauri disks \citep{pont10,naji,fed}. From the theoretical side, several attempts have been made to explain the formation and survival of molecules (especially water) in disks and their abundance during the early phases of planet formation \citep[e.g.][]{ciesla,glas,bb09,vasy}. How does the warm molecular gas composition depend on stellar+disk properties, and how may the interplay and evolution of the different components enable (instead of hinder) the availability of essential molecules for planet formation? The most recent outburst in EX Lupi has provided an extraordinary opportunity to address this compelling questions. While it is difficult to identify one factor that is the main responsible for the diversity in molecular emission of classical T Tauri systems, due to the many properties that vary from system to system, in the case of EX Lupi we can study a single system where only \textit{one} parameter (the stellar+accretion luminosity) dramatically increases, and in a timescale that is feasible for us to monitor. Crystal formation in outburst from amorphous dust particles was recently shown by \citet{abra09} comparing \textit{Spitzer} spectra of EX Lupi, suggesting that episodic increases in disk temperature during outbursts may contribute significantly to the formation of some ingredients that are now present in our Solar System. Using the same spectra, but focussing on the gas, we show here how H$_{2}$O (water) emission varies in outburst together with other molecules like OH (hydroxyl), H \textsc{i} and H$_{2}$ (atomic and molecular hydrogen), and organics like C$_{2}$H$_{2}$ (acetylene), HCN (hydrogen cyanide), and CO$_{2}$ (carbon dioxide), giving the opportunity to investigate the conditions of the emitting gas and the availability and survival of these important molecules during the unstable early phases in disk evolution.

\section{OBSERVATIONS} \label{sec: obs}
\subsection{\textit{Spitzer} IRS Archival Spectra}
We collected all archival spectra of EX Lupi obtained with the Infrared Spectrograph (IRS) on the \textit{Spitzer} satellite \citep{houck04}. High-resolution spectroscopy with the IRS is available in two different modules with a resolving power R $\approx$ 600: the short-high (SH) module in the wavelength range 9.9--19.6  $\mu$m and the long-high (LH) module between 18.7 and 37 $\mu$m. The IRS low-resolution modules instead cover the range 5.2--38 $\mu$m with a resolving power R $\approx$ 60--120. The IRS observations of EX Lupi span the epoch surrounding its latest outburst in 2008. In a quiescent phase before the outburst two different spectra were acquired: in 2004 (PI: N. Evans, hereafter E04) and in 2005 (PI: G. Stringfellow, hereafter S05), with both low- and high-resolution modules. The outburst phase was then observed in April 2008 (PI: P. \'Abrah\'am, hereafter A08) and May 2008 (PI: J. Carr, hereafter C08) with the high-resolution modules, while post-outburst spectra during a new quiescent phase were taken in October 2008 and April 2009 (PI: P. \'Abrah\'am, hereafter A09) only with low resolution. In Table \ref{tab:arch} all the \textit{Spitzer} IRS observations of EX Lupi are listed with their relevant observational parameters. Since the resolution of post-outburst spectra (A09) is too low to detect the large majority of the gas emission lines studied in this work, we restricted our investigation to the high-resolution spectra taken before and during outburst. More precisely, we decided to use in the present analysis only the spectra closer in time to the onset of the 2008 outburst, to study the changes in emission that can be related to it: in the quiescent phase the S05 spectrum and in outburst the A08 spectrum. The investigation of variations/correlations in emission over a broader range of timescales during a quiescent period (comparing E04 and S05) and short-term changes in outburst (comparing A08 and C08) may be addressed in a follow-up paper.

\begin{figure*}
\includegraphics[width=0.95\textwidth]{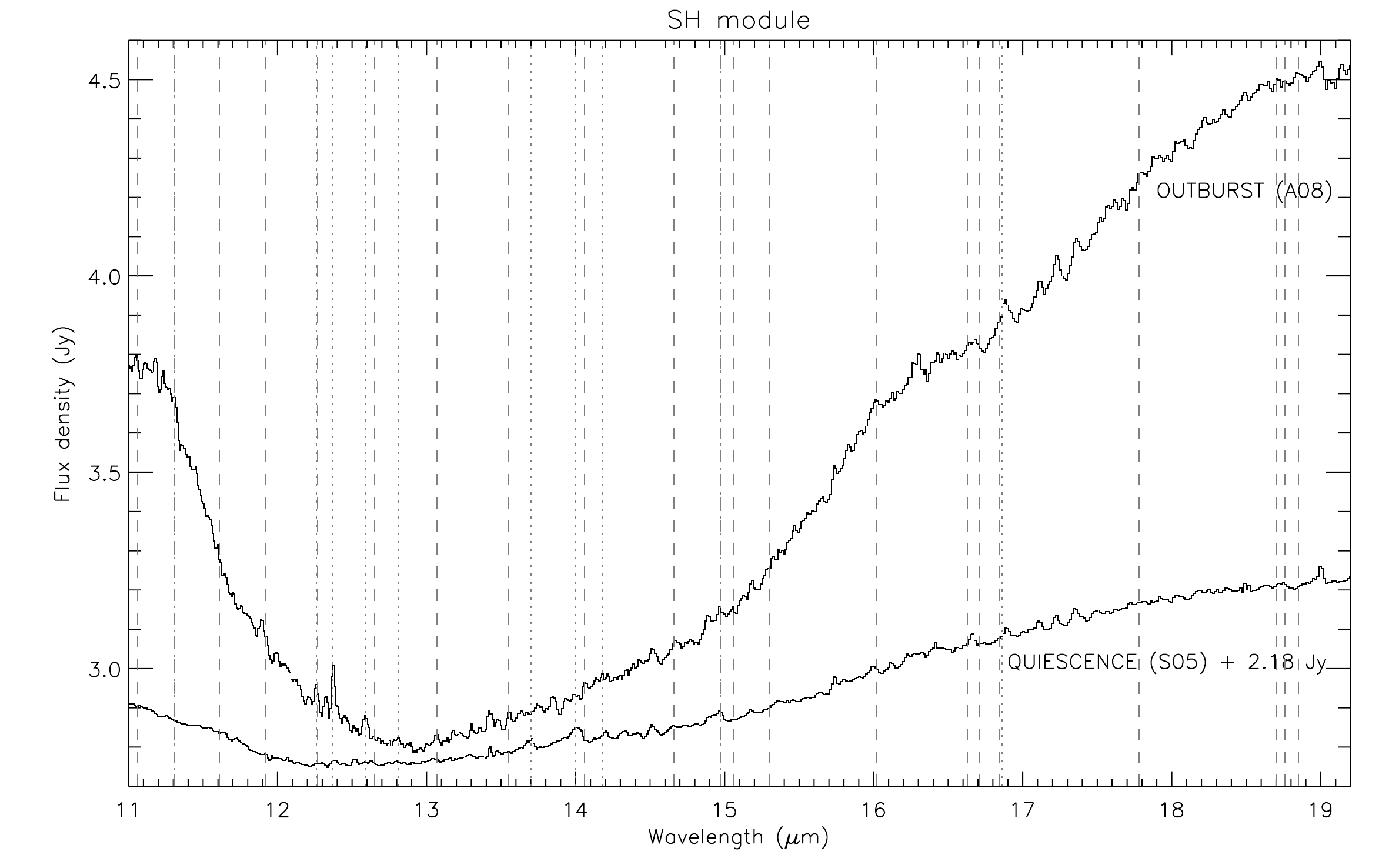} 
\includegraphics[width=0.95\textwidth]{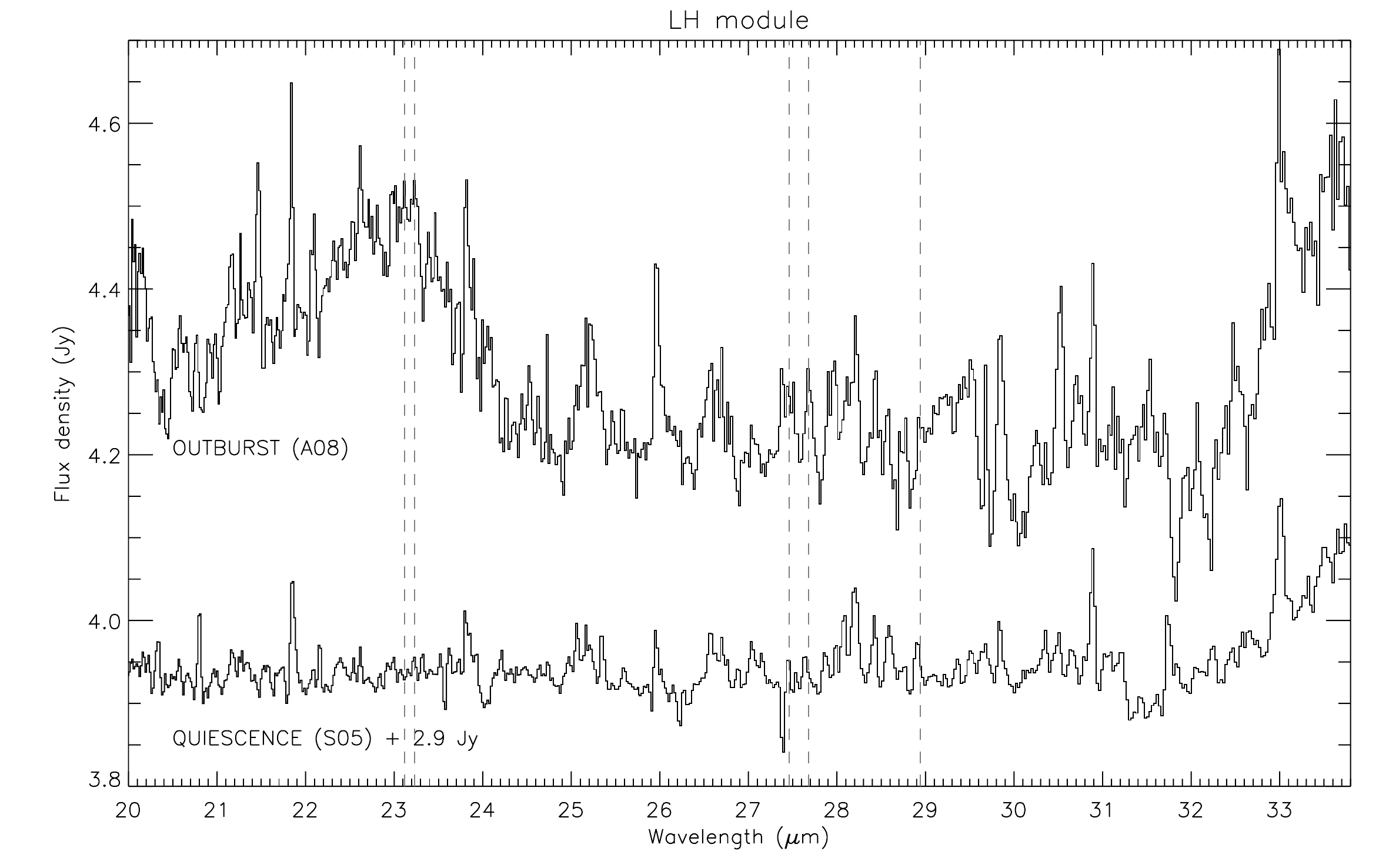} 
\caption{Comparison between the quiescent (S05) and the outburst (A08) phases in the \textit{Spitzer} spectra of EX Lupi: the SH module is shown in the top and the LH module in the bottom. The quiescent spectrum is shifted for comparison in both plots. OH transitions are marked with dashed lines, while other species (C$_{2}$H$_{2}$, HCN, CO$_{2}$, H \textsc{i}, H$_{2}$, [Ne \textsc{ii}]) are marked with dotted lines. All unmarked lines are identified as water rotational transitions.}
\label{fig:spec_comp}
\end{figure*}

\subsection{Data Reduction} \label{sec: obs2}
We re-reduced the IRS spectra starting from the \textit{Spitzer} Science Center S18.7.0 products and applying the latest version of the pipeline provided by the ``Cores to Disks" \textit{Spitzer} legacy program, described in Lahuis et al. (2006). This pipeline was developed to reduce IRS pointed observations and reach high sensitivity particularly when observing faint objects. The extraction of one-dimensional spectra from the images is done using  an optimal extraction, where the IRS point spread function, defined using sky-corrected high signal-to-noise (S/N) calibrators, is fitted on good pixels only (i.e. excluding known bad/hot pixels). Dither positions are first combined and then the spectrum is extracted, which gives the best resultant S/N. Moreover, this method provides an estimate of the local sky contribution, which was separately observed with \textit{Spitzer} only in the case of A08.

The EX Lupi spectra, like all IRS spectra of bright point sources, are affected by systematics that reduce the maximum S/N achievable with long exposures (e.g. uncertainties in pointing correction, flux-dependent calibration, etc.). These uncertainties are difficult to account for and are not included in the errors on individual pixels propagated through the reduction pipeline. In our analysis we chose to derive our own estimate of the noise as follows. We measure in each spectrum the dispersion of pixels from a baseline (first-order polynomial) fitted to the continuum. Unfortunately, given the spectral resolution of \textit{Spitzer}, a clean continuum is difficult to find and often we can constrain only a pseudo-continuum given by weak emission lines. We use the HITRAN database to avoid ranges where the emission from obvious molecules is likely to be strong (see Section \ref{sec: anal1}). The pixel dispersion with reference to the baseline is then checked for gaussianity with the Shapiro-Wilk test \citep{shap}. We consider four spectral ranges in SH (11.95--12.2, 15.44--15.6, 17.45--17.7 and 19.05--19.2 $\mu$m) and four in LH (24.15--24.55, 25.45--25.85, 30--30.2 and 31.3--31.65 $\mu$m) as an attempt to account for the variations of noise with wavelength. We discard the rms estimated in a given range if the probability from the Shapiro-Wilk test that the data are consistent with having been drawn from a Gaussian is less than 10\%. As a result, in quiescence (S05) we find an rms of $\sim$3 and $\sim$9 mJy in SH and LH respectively, while in outburst (A08) we find $\sim$8 and $\sim$20 mJy respectively. We assume these estimates as the flux uncertainty on individual pixels in the two IRS modules separately, and we use them in the derivation of line flux errors as described in Section \ref{sec: anal2}. On the one hand these should be regarded as conservative estimates as the molecular emission in the chosen ranges is \textit{expected} to be small, but weak features (as well as chemical species that are not considered) may still contribute to the pixel dispersion and thus artificially increase the noise. On the other hand, if high systematics strongly affects \textit{Spitzer} spectra our estimates might still be underestimating the true uncertainty \citep[cf.][]{sal11,cn11}. However, we have some indications that our estimates of the noise are conservative rather than optimistic (see Sections \ref{sec: anal2} and \ref{sec: res1}).

\begin{figure}
\includegraphics[width=0.47\textwidth]{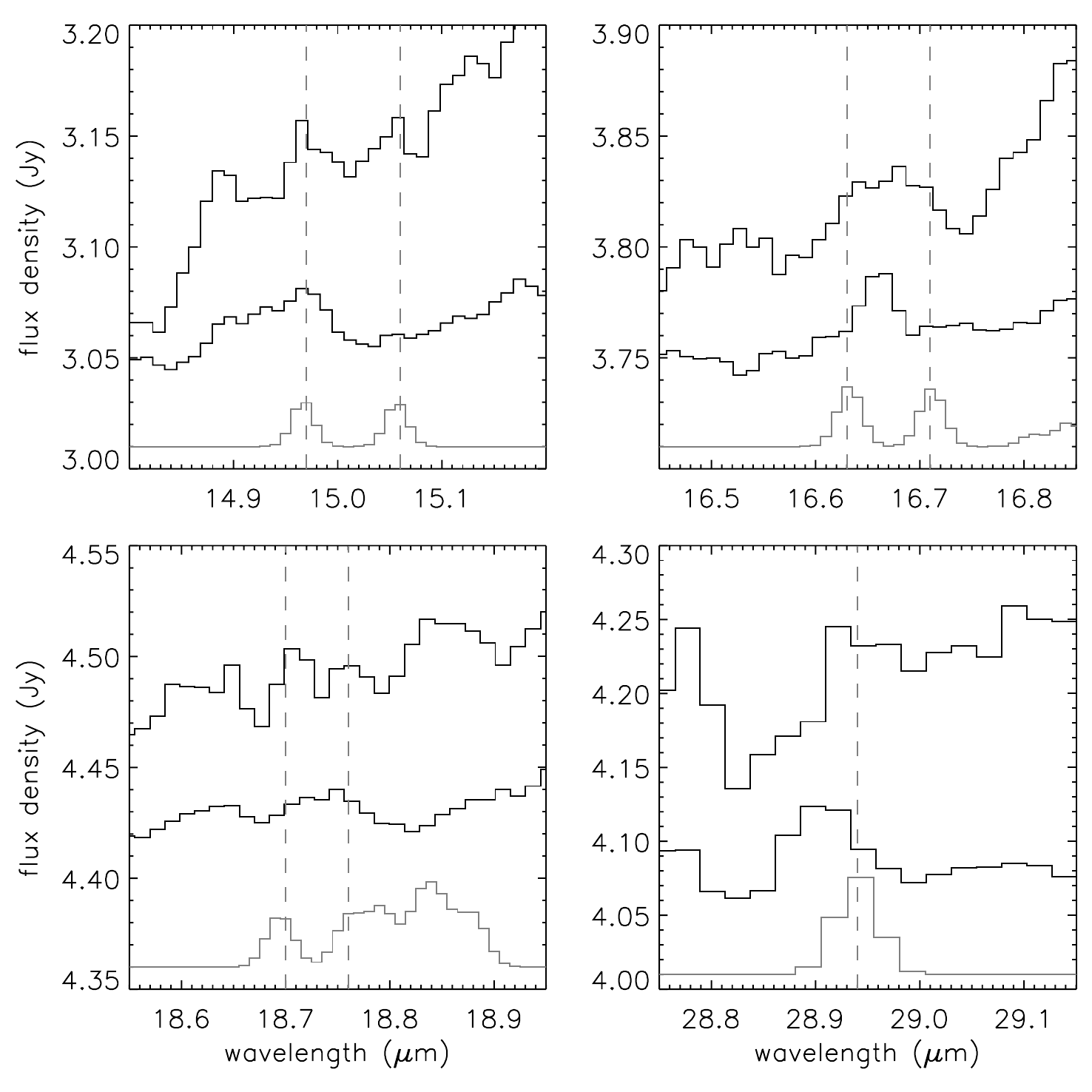} 
\caption{Cross-ladder OH transitions detected in EX Lupi in outburst. Each plot shows, from top to bottom, the outburst A08 spectrum, the quiescent S05 spectrum (shifted in flux for comparison), and an LTE model for OH with $T_{\rm ex} \sim 600$ K and $N_{\rm mol} \sim$ 3 $\times 10^{17}$ cm$^{-2}$. In the OH model, which includes both intra-ladder and cross-ladder transitions, the position of cross-ladder transitions is indicated with vertical dashed lines. In each plot, observed emission lines that do not match the OH model shown in the bottom are attributed to H$_{2}$O, apart from the 14.97 $\mu$m feature in quiescence due to CO$_{2}$.}
\label{fig:cl_trans}
\end{figure}

\begin{deluxetable*}{l c c c c c}
\tabletypesize{\small}
\tablewidth{500pt}
\tablecaption{\label{tab:oh}OH Detections in EX Lupi.}
\tablehead{\colhead{$\lambda$ ($\mu$m)} & \colhead{Transitions} & \colhead{$E_{u}$ (K)} & \colhead{$A_{ul}$ (s$^{-1}$)} & \colhead{S05} & \colhead{A08}}
\startdata
11.06 & 1/2 (57/2 $\rightarrow$ 55/2), 3/2 (59/2 $\rightarrow$ 57/2) & 21,300 & 1300 & - & $\surd$ \\
12.65 & 1/2 (47/2 $\rightarrow$ 45/2), 3/2 (49/2 $\rightarrow$ 47/2) & 15,100 & 850 & $\surd$ & - \\
13.07 & 1/2 (45/2 $\rightarrow$ 43/2), 3/2 (47/2 $\rightarrow$ 45/2) & 13,900 & 780 & - & $\surd$ \\
13.55  & 1/2 (43/2 $\rightarrow$ 41/2), 3/2 (45/2 $\rightarrow$ 43/2) & 12,800 & 690 & - & $\surd$ \\
14.07  & 1/2 (41/2 $\rightarrow$ 39/2), 3/2 (43/2 $\rightarrow$ 41/2) & 11,800 & 620 & - & $\surd$ \\
14.65  & 1/2 (39/2 $\rightarrow$ 37/2), 3/2 (41/2 $\rightarrow$ 39/2) & 10,800 & 540 & $\surd$ & $\surd$  \\
15.00  & 1/2 (17/2) $\rightarrow$ 3/2 (15/2) & 2400 & 0.2 & - & $\surd$ \\
16.66  & 1/2 (15/2) $\rightarrow$ 3/2 (13/2) & 2000 & 0.2 & - & $\surd$ \\
18.73  & 1/2 (13/2) $\rightarrow$ 3/2 (11/2) & 1550 & 0.2 & - & $\surd$ \\
18.85  & 1/2 (29/2 $\rightarrow$ 27/2) & 6250 & 250 & - & $\surd$  \\
23.10  & 3/2 (25/2 $\rightarrow$ 23/2) & 4100  & 130 & - & $\surd$  \\
23.22 & 1/2 (23/2 $\rightarrow$ 21/2) & 4150 & 130 & ? & $\surd$ \\
27.42  & 3/2 (21/2 $\rightarrow$ 19/2) & 2900 & 80 & - & $\surd$  \\
27.67  & 1/2 (19/2 $\rightarrow$ 17/2) & 2960 & 80 & ? & $\surd$  \\
28.94  & 1/2 (7/2) $\rightarrow$ 3/2 (5/2) & 620 & 0.1 & ? & $\surd$  
\enddata
\tablecomments{At the spectral resolution of the \textit{Spitzer} IRS, all observed OH lines are blends of doublets of intra-ladder or cross-ladder transitions belonging to the two pure rotational ladders $^{2}\Pi_{3/2}$ and $^{2}\Pi_{1/2}$ in the ground vibrational state. In this table we provide in parentheses the upper and lower \textit{J}-level of each doublet. Cross-ladder doublets are reported as a single feature even when two lines can be distinguished (see Figure \ref{fig:cl_trans}), and in the rotation diagram analysis we use the unified flux (Section \ref{sec: res2}). The wavelengths we report are approximated at the center of each feature. Molecular data are taken from the HITRAN database. The $A_{ul}$ reported here are the sums over the individual transitions blended in each observed line. The last two columns on the right indicate the detection in the quiescent (S05) and outburst (A08) spectra (S/N $>$ 2$\sigma$). A question mark shows lines that are not unambiguously identified as OH.}
\end{deluxetable*}

\section{SPECTRAL LINE ANALYSIS} \label{sec: anal}
\subsection{Molecular and Atomic Emission} \label{sec: anal1}
The identification of molecular emission features in the \textit{Spitzer} observations of EX Lupi is based on the comparison between the observed spectra and synthetic models generated using the HITRAN 2008 database \citep{hitran}. The MIR spectra of EX Lupi are dominated by a forest of water emission lines\footnote{In the text we distinguish ``lines" from ``transitions". The former refers to the unresolved emission features observed in the spectra, while the latter to the individual molecular transitions that produce the unresolved observed features.}, placing this source within the sample of ``wet" T Tauri disks recently revealed by \textit{Spitzer} \citep{cn08,cn11,sal08,pont10}. H$_{2}$O lines are detected in both the quiescent and the outburst phases of EX Lupi and their measured flux with respect to the continuum increases in outburst over the entire spectral range covered by \textit{Spitzer}, on top of the continuum level that is $\sim$4 times higher than in quiescence (see Figure \ref{fig:spec_comp}). The water vapor emission is composed of unresolved pure rotational transitions in the ground vibrational state with upper level energies ($E_{u}$) of a few 1000 K. The complex excitation structure of water makes its observational study a real challenge, especially at low spectral resolution. Even small spectral ranges in the infrared can be finely populated by rotational transitions from different $E_{u}$, which are resolved only with a spectral resolution much higher than that provided by \textit{Spitzer} \citep[see e.g. Figure 5 in][]{pont10}. The criticality of such a limitation on the derivation of the emitting gas properties will be addressed in Sections \ref{sec: model} and \ref{sec: res}.

In addition to water, the MIR spectra of EX Lupi reveal a variety of other molecules and atoms with a noticeable difference between the two activity phases of the star. In quiescence the rovibrational Q-branches of C$_{2}$H$_{2}$ at $\sim$13.7 $\mu$m, HCN at $\sim$14 $\mu$m, and CO$_{2}$ at $\sim$15 $\mu$m are clearly detected, while a weak contribution from OH is identified. In outburst organic molecules are no longer seen, while strong H \textsc{i} 7--6, H \textsc{i} 14--9, H$_{2}$ S(2), and several new OH lines appear and increase the confusion with water emission, especially in the SH module.
The OH emission consists of rotational transitions in the ground vibrational state with $E_{u}$ from $\sim$600 up to $\sim$20,000 K and is detected in outburst throughout the entire spectrum. These transitions are mainly intra-ladder transitions belonging either to the $^{2}\Pi_{3/2}$ or the $^{2}\Pi_{1/2}$ pure rotational ladders, but several cross-ladders transitions are also identified, indicating a remarkable population of intermediate and low $E_{u}$ (see Figure \ref{fig:cl_trans} and Table \ref{tab:oh}). To our knowledge, this is the first time that cross-ladder transitions are identified in disks, especially those with $E_{u}\sim$ 1000-2000 K. \citet{tap}, in the HH211 outflow, were the first to ever report the detection of OH transitions from levels as high as $\sim$28,200 K, and they also identified two lower-$E_{u}$ cross-ladder transitions (one of which, at 28.94 $\mu$m, is also detected by us in the EX Lupi outburst spectrum). In EX Lupi, additional OH lines from $E_{u}$ as high as those found by \citet{tap} might be present in the outburst spectrum, but their detection is hindered by the strong silicate feature at $\sim$10 $\mu$m. We will come back to the analysis of OH emission, and the importance of the simultaneous detection of low- and high-$E_{u}$ lines in outburst, in Section \ref{sec: res2}.

In \textit{Spitzer} spectra all the chemical species mentioned above are blended with H$_{2}$O lines. This, in addition to the above mentioned blending of transitions from different energy levels, hinders not only the molecular line identification but also the derivation of gas properties. Excluding the organic molecules, whose Q-branches are broader than the gas-phase water lines, it can be misleading to identify molecules from only 1-2 emission lines observed where their emission is expected to be, unless such lines prove to be clearly separated by other species. The case of OH is particularly good as an example. In outburst we can identify all the expected lines from the two rotational ladders in the ground level, thus the molecule is certainly detected. In quiescence, instead, only a few lines are unambiguously identified, at 12.65 and 14.65 $\mu$m, which are the most spectrally separated by H$_{2}$O lines. All other tentative OH detections are confused with other molecules. The case of OH lines at longer wavelengths (the LH module) is particularly difficult to judge, especially because of the unconstrained H$_{2}$O emission (see Section \ref{sec: res1}). However, consideration of reasonable models for the H$_{2}$O emission suggests that the data in LH in the S05 spectrum can be explained as H$_{2}$O emission alone. Therefore the (partial) detection of OH in quiescence is based on two lines in the SH module, while in the LH module it is highly dubious.

Other chemical species that have been identified in the MIR in other T Tauri stars may additionally contribute to the emission observed in EX Lupi \citep[see e.g.][]{naji}. However, in the present paper we focus our analysis on H$_{2}$O and OH, which largely dominate the MIR emission that we observe in EX Lupi. We assume that the MIR emission from these two molecules originates in the innermost disk atmosphere ($\sim$1--10 AU from the central star), as proposed for other systems in similar recent studies \citep{cn08,cn11,pont10,sal11} and confirmed by higher-resolution observations for a few objects \citep{pont10b,fed}.

\begin{figure}
\includegraphics[width=0.47\textwidth]{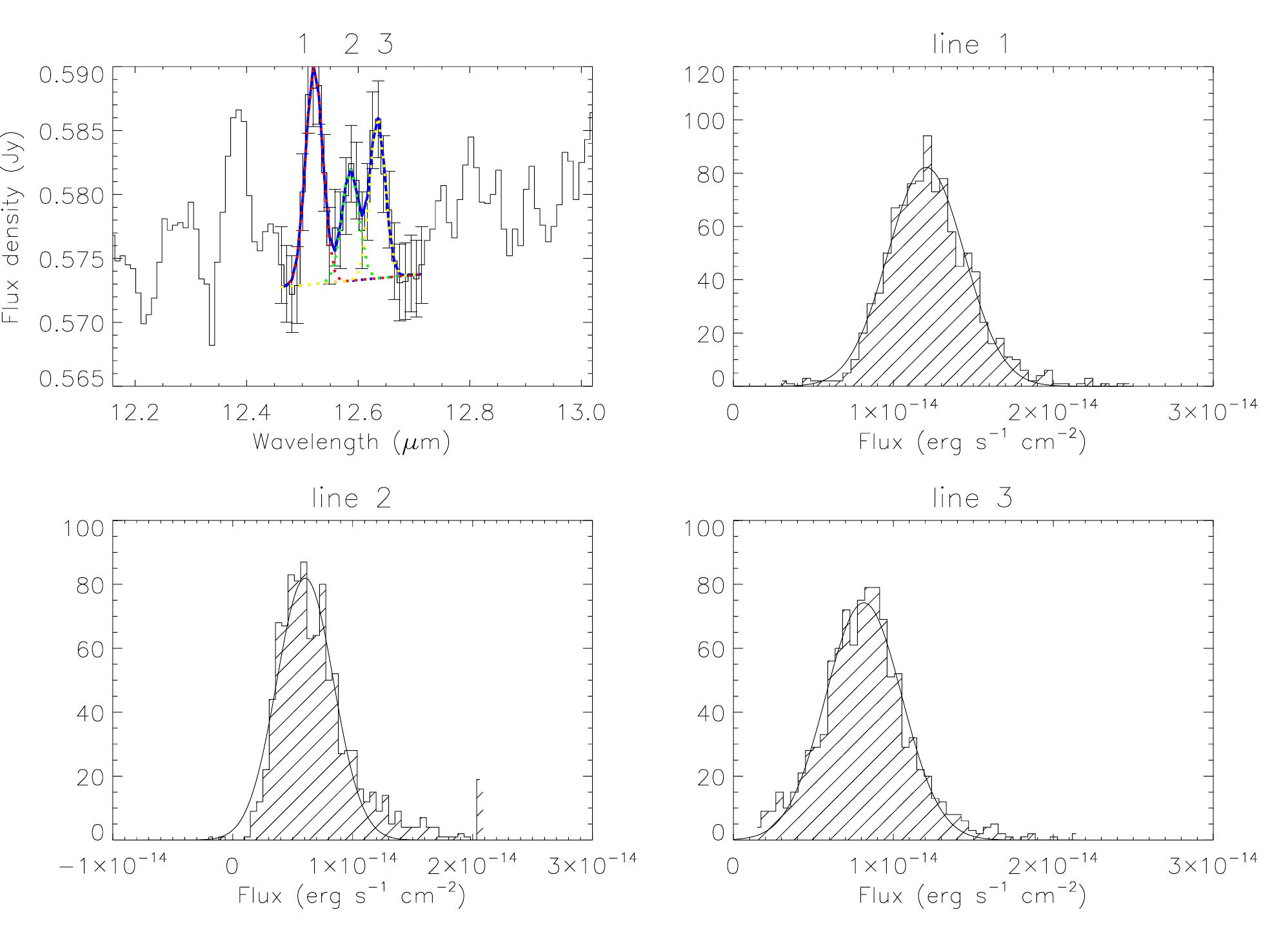} 
\caption{Measure of line flux and error for emission lines in the S05 spectrum. In the first plot (top, left) the best-fit curve (first-order polynomial + three Gaussians) is overplotted to the original data, where the error on individual pixels is set to the measured rms dispersion (see Section \ref{sec: obs2}). The $\chi_{red} ^2$ of the best fit is 0.2. Such a low value suggests that our rms estimate is conservative. The error on each line flux is set to a robust standard deviation of the distribution of 1000 repetitions of the flux measure (see Section \ref{sec: anal2}). The first two lines are H$_{2}$O lines, while the third is OH.}
\label{fig:mc}
\end{figure}

\begin{deluxetable*}{c c l c l c}
\tabletypesize{\small}
\tablewidth{500pt}
\tablecaption{\label{tab:fluxes} Flux Estimates of Emission Lines in the IRS \textit{Spitzer} Spectra of EX Lupi.}
\tablehead{\colhead{Spectral Range ($\mu$m)} & Line ($\mu$m) & \multicolumn{2}{c}{Quiescence (S05)} & \multicolumn{2}{c}{Outburst (A08)} }
\tablecolumns{6}
\startdata
11.00--11.10 & 11.06 & ... &  ... &  OH & 2.63 $\pm$ 0.51 \\
12.22--12.50 & 12.28 & OH+H$_{2}$O  & 1.22 $\pm$ 0.55 &  H$_{2}$+OH+H$_{2}$O & 2.87 $\pm$ 0.43 \\
12.22--12.50 & 12.37 & H$_{2}$O & 1.51 $\pm$ 0.46 &  H \textsc{i}+H$_{2}$O & 6.78 $\pm$ 0.59 \\
12.46--12.72 & 12.51 & H$_{2}$O   & 1.23 $\pm$ 0.25 & H$_{2}$O & ... \\
12.46--12.72 & 12.58 & H$_{2}$O  & 0.62 $\pm$ 0.23  & H \textsc{i}+H$_{2}$O & 3.71 $\pm$ 0.61 \\
12.46--12.72 & 12.64 & OH & 0.82 $\pm$ 0.26  &  OH & ... \\
12.67--12.90 & 12.81 & [Ne \begin{scriptsize}II\end{scriptsize}]+OH & 0.66 $\pm$ 0.28  &  [Ne \begin{scriptsize}II\end{scriptsize}]+OH & 2.59 $\pm$ 1.13\\ 
12.95--13.15 & 13.06 & H$_{2}$O+OH  &  0.73 $\pm$ 0.38 &  OH & 1.33 $\pm$ 0.51 \\
13.40--13.60 & 13.43 & H$_{2}$O  & 0.90 $\pm$ 0.14  & H$_{2}$O & 3.27 $\pm$ 0.53 \\
13.40--13.60 & 13.49 & H$_{2}$O   & 0.39 $\pm$ 0.16  & H$_{2}$O & 0.86 $\pm$ 0.31 \\
13.40--13.60 & 13.55 & OH  &  ... &  OH & 1.51 $\pm$ 0.44 \\
13.50--13.80 & 13.70 & C$_{2}$H$_{2}$+H$_{2}$O  & 2.76 $\pm$ 0.29 & ... &  ...  \\
13.75--14.12 & 14.00 & HCN+H$_{2}$O  & 7.48 $\pm$ 0.63 & ... &  ... \\
13.95--14.30 & 14.06 & OH  &  ... &  OH & 1.22 $\pm$ 0.37 \\
13.95--14.30 & 14.20 & H$_{2}$O  & 1.38 $\pm$ 0.30 & H$_{2}$O & 1.71 $\pm$ 0.56 \\
14.43--14.61 & 14.51 & H$_{2}$O  & 1.87 $\pm$ 0.20  & H$_{2}$O & 2.08 $\pm$ 0.46 \\
14.55--14.85 & 14.65 & OH  & 1.10 $\pm$ 0.39 &  OH & 3.51 $\pm$ 1.19 \\
14.80--15.05 & 14.95 & CO$_{2}$+H$_{2}$O & 3.20 $\pm$ 0.29 & H$_{2}$O & 3.36 $\pm$ 0.77 \\
14.93--15.10 & 15.00 & ... & ... &  OH & 1.35 $\pm$ 0.66 \\
15.67--15.87 & 15.74 & H$_{2}$O  & 1.48 $\pm$ 0.17  & H$_{2}$O & 2.04 $\pm$ 0.45  \\
15.85--16.10 & 16.00 & H$_{2}$O  & 1.43 $\pm$ 0.22  & H$_{2}$O+OH & 4.34 $\pm$ 0.95 \\
16.05--16.18 & 16.11 & H$_{2}$O  & 0.85 $\pm$ 0.19  & H$_{2}$O & ...  \\
16.53--16.80 & 16.66 & H$_{2}$O  & 1.47 $\pm$ 0.18  & H$_{2}$O+OH & 2.87 $\pm$ 0.76  \\
16.73--16.99 & 16.89 & H$_{2}$O  & 1.34 $\pm$ 0.22  & H$_{2}$O+OH & 5.76 $\pm$ 0.84  \\
17.03--17.19 & 17.11 & H$_{2}$O  & 1.01 $\pm$ 0.14  & H$_{2}$O & 1.61 $\pm$ 0.66  \\
17.14--17.31 & 17.22 & H$_{2}$O  & 1.26 $\pm$ 0.15  & H$_{2}$O & 3.36 $\pm$ 0.60  \\
17.28--17.46 & 17.36 & H$_{2}$O  & 2.14 $\pm$ 0.28  & H$_{2}$O & 2.70 $\pm$ 0.39 \\
18.65--18.80 & 18.73 & ...  &  ...  &  OH & 1.45 $\pm$ 0.55 \\
18.73--19.11 & 18.85 & ...  &  ...  &  OH & 2.17 $\pm$ 0.61 \\
18.73--19.11 & 19.00 & H$_{2}$O  & 1.24 $\pm$ 0.11 & H$_{2}$O & 3.37 $\pm$ 0.97  \\
20.68--20.90 & 20.79 & H$_{2}$O  & 2.85 $\pm$ 0.28  & H$_{2}$O & 2.78 $\pm$ 0.74  \\
21.70--22.04 & 21.84 & H$_{2}$O  & 5.51 $\pm$ 0.44  & H$_{2}$O & 6.52 $\pm$ 0.57  \\
22.82--23.37 & 23.10 & ...  & ...  &  OH & 7.26 $\pm$ 2.87  \\
22.82--23.37 & 23.22 & H$_{2}$O & 0.50 $\pm$ 0.19  &  OH  & 5.64 $\pm$ 1.64  \\
27.20--27.88 & 27.42 & ...  & ...  &  OH & 6.43 $\pm$ 1.43  \\
27.20--27.88 & 27.67 & H$_{2}$O  & 1.21 $\pm$ 0.35  &  OH & 4.58 $\pm$ 0.82  \\
28.81--29.07 & 28.94 & H$_{2}$O  & 1.23 $\pm$ 0.22 & OH  & 1.45 $\pm$ 0.56  \\
29.72--30.10 & 29.84 & H$_{2}$O  & 1.66 $\pm$ 0.27  & H$_{2}$O & 9.39 $\pm$ 1.19  \\
30.40--31.05 & 30.49 & H$_{2}$O & 1.05 $\pm$ 0.26 & H$_{2}$O & 6.88 $\pm$ 0.92  \\
30.40--31.05 & 30.88 & H$_{2}$O  & 3.12 $\pm$ 0.25  & H$_{2}$O & 4.61 $\pm$ 0.51  \\
32.78--33.40 & 33.00 & H$_{2}$O  & 5.03 $\pm$ 0.36 & H$_{2}$O & 8.62 $\pm$ 0.82  
\enddata
\tablecomments{We give the spectral range of the observed emission features where we apply our method to derive line fluxes and errors (see Section \ref{sec: anal2}). For each range we indicate the identified species most contributing to the measured flux. In the cases where it was possible to distinguish different lines within a feature (e.g. see Figure \ref{fig:mc}), we report the components separately. The line flux is given in units of $10^{-14}$ erg s$^{-1}$ cm$^{-2}$. All fluxes reported in the table have S/N $>$ 2$\sigma$. Where an entry in the table is omitted, emission lines are not clearly identified and/or the measured flux is consistent with zero.}
\end{deluxetable*}

\subsection{Estimation of Emission Line Fluxes and Errors} \label{sec: anal2}
Given the need for reliable tracers to probe the emission variability, we chose to use only the most trustable emission lines and to measure their flux and error using a robust method based on our rms estimate. We restrict the analysis to strong and/or isolated emission lines which we believe are the least affected by artifacts. For instance, the lines in spectral ranges where different spectral orders overlap, which are affected by drop of signal, and those where the continuum or pseudo-continuum is more uncertain are excluded from the analysis. In Table \ref{tab:fluxes} we list the emission lines considered in this paper, indicating the spectral range of each observed feature (including the nearby continuum), the chemical species identified as contributing to the emission, and the flux measured in quiescence and outburst. 

Our method partly resembles techniques previously utilized for the measure of line fluxes in \textit{Spitzer} spectra \citep[e.g.][]{pont10,naji}, and for the estimate of flux uncertainties \citep{pasc08}. We developed our method in IDL building on publicly available routines as follows. For each observed emission line we select a spectral range including at least three pixels on each side for the continuum (typically a few tenths of a micron). Where a continuum cannot be found in between two or more lines because of their vicinity, we fit them together. We fit the data using a least-squares Levenberg-Marquardt algorithm \citep{lev-mar} with a baseline plus one Gaussian function for each line considered, with width and area as free parameters. Since we fit the continuum locally, we always assume a first-order polynomial for the baseline even though a higher order could in some cases slightly improve the fit. Each line flux is computed as the area of the locally continuum-subtracted best-fit Gaussian. The observed emission line(s) must be well represented by one (or more) Gaussian functions, and we set an acceptance threshold to the goodness of the fit to $\chi^{2}_{red} \lesssim 1.5$. In most cases the $\chi^{2}_{red}$ is much lower than that limit, supporting that our noise estimate is indeed conservative (see e.g. Figure \ref{fig:mc}). This method provides the possibility to measure the individual line fluxes even in the case of complexes composed of several blended lines from the same or different chemical species (an example is shown in Figure \ref{fig:mc}), unless the different lines are overlapping too much to be distinguished, as it is often the case in the LH module. This approach is not used for the broad emission features from C$_{2}$H$_{2}$, HCN, and CO$_{2}$, because the shape of the Q-branch, which is at least partially resolved by the IRS, is not Gaussian. We therefore fit only a baseline continuum to nearby regions as described above and then integrate the flux below the continuum-subtracted data. 

The 1-$\sigma$ error on each line flux is then estimated using the following Monte-Carlo based approach. In each of the spectral regions considered above, we add normally distributed noise to the observed flux of individual pixels, that we take as the mean. As the noise to be added we use the rms calculated from the supposed line-free regions (Section \ref{sec: obs2}). The procedure is repeated 1000 times and we measure the line fluxes from each realization exactly as described above. The error on the line flux is then set to a robust estimate of the standard deviation of the distribution of its 1000 realizations (see Figure \ref{fig:mc}). With this method we take into account the noise on single pixels (given by our estimate of the rms) as well as the uncertainty on the local fit to the continuum. 
\\

\section{CONSTRAINING GAS PROPERTIES} \label{sec: model}
\subsection{An LTE Approach} \label{sec: model1}
A simple approximation utilized in observational astrochemistry is to assume the emitting gas to be in local thermal equilibrium. In LTE the population of the excited levels simply follows the Boltzmann distribution, i.e. all levels are thermalized to the kinetic temperature ($T_{\rm k}$) of the gas. A transition between two levels can be described as to be in LTE when collisions dominate the excitation/de-excitation, i.e. when the density of the ambient gas is higher than the critical density of the transition ($n_{\rm crit} = A_{ul} / C_{ul}$, where $A_{ul}$ is the Einstein-A coefficient and $C_{ul}$ the collision rate). If all transitions are in LTE, the kinetic temperature and the column density of the gas can be derived from the observed flux of spectrally-resolved transitions. The techniques that are most frequently used to derive gas properties are based on the rotation diagram analysis \citep[e.g.][hereafter GL99]{GL99} or the direct fit of transition intensities \citep[e.g.][hereafter ME09]{me09}. In the extremely diverse conditions found in astrophysical environments, however, it is often the case that the ambient density is insufficient to thermalize some (or all) levels and the LTE description falls short of deriving the real gas properties. ME09 showed how the LTE approximation is inappropriate for circumstellar disks, as the wide range of thermodynamic conditions found in the inner $\sim10$ AU of disks (gas temperatures ranging from $\sim100$ to a few 1000 K, and densities from $\sim 10^3$ to $10^{16}$ cm$^{-3}$) results in emission produced by a mixture of thermalized and non-thermalized levels. GL99 showed how the rotation diagram analysis can help in distinguishing the effects of optical depth and non-thermal excitation on the typical straight-line behavior of optically thin lines in LTE. However, for this type of analysis spectrally resolved observations are required, to separate different energy levels. With the spectral resolution provided by \textit{Spitzer} this is possible in the case of OH, but not water. The implementation of such a technique in the case of OH emission will be presented in Section \ref{sec: res2}. Now we focus on the method used to derive constraints on the excitation properties of the unresolved and blended water emission.

Non-LTE models that perform line radiative transfer and account for the geometrical and physical structure of disks have been explored to some extent \citep[e.g.][]{pavly}, but they are not yet viable alternatives for constraining water MIR line emission because of the incompleteness and uncertainty of collision data \citep[see][]{vdt11}. This is why the simple LTE approximation is still the first approach utilized when modelling the molecular emission observed in \textit{Spitzer} spectra \citep{cn08,cn11,sal11}. We therefore developed our own simple single-slab LTE model to constrain the molecular emission in EX Lupi. The model is explained in detail in Appendix \ref{app: model}, while here we provide only a short description. The free parameters of the model are the excitation temperature $T_{\rm ex}$, the column density $N_{\rm mol}$, and the emitting area $A$. In the fit results we will report the value of the radius, $r_{\rm p}$ (projected), of the circular area that is equivalent to $A$. The excitation and relative strength of different transitions are sensitive to $T_{\rm ex}$ and $N_{\rm mol}$, while $A$ only has the effect of a scaling factor common to all line fluxes. Observed and model spectra are compared using the individual fluxes of blended emission lines (given in Table \ref{tab:fluxes}), and a $\chi^2$ test is performed minimizing over the difference between observed and model line fluxes as an ensemble. Despite the fact that the model parameters are sensitive to different properties and that we account for line optical depth effects, the fit results are degenerate in $T_{\rm ex}$, $N_{\rm mol}$, and $A$. In general, for a given emitting area the fit is degenerate with higher $T_{\rm ex}$ for lower $N_{\rm mol}$ (and vice versa, see e.g. results in Section \ref{sec: res1}). In the extreme optically thick regime ($\tau\gg1$) the line flux depends on the product $A \times T_{\rm ex}$, thus only those can be derived while the model is insensitive to changes of $N_{\rm mol}$. In the extreme optically thin case ($\tau\ll1$), instead, the line flux is proportional to $A \times N_{\rm mol} \times T_{\rm ex}$, and we can reduce again to two parameters: the temperature and the number of emitting molecules = $A \times N_{\rm mol}$, yet still suffering of the degeneracy between $T_{\rm ex}$ and $N_{\rm mol}$. Ideally, three emission lines that are differently sensitive to variations in $T_{\rm ex}$ and $N_{\rm mol}$ should be enough to constrain all model parameters together. Water MIR transitions do cover a wide range of opacities and therefore an appropriate selection of them could serve the scope. However, in practice their different opacity cannot be retrieved unambiguously from the blended lines they produce in the \textit{Spitzer} spectra and the degeneracy between model parameters is somewhat unavoidable.
\\

\subsection{Fitting Spitzer Spectra of H$_{2}$O Emission with LTE Models} \label{sec: model2}
Fitting the MIR unresolved water emission in EX Lupi using LTE single-slab models is challenging. One must keep in mind that such simplified models do not account for any disk geometry, structure, nor temperature/column-density profiles, which are complex \citep[see e.g.][]{ber07} and in the case of EX Lupi might also change from quiescence to outburst. In addition to that, flux contamination from different molecules can also affect the fits, unless different contributions can be individually constrained and removed (which usually require many ``clean" lines of the same species and a reliable modelling). In fact, we found that in EX Lupi it is impossible to get a good fit over the wide spectral range covered by the \textit{Spitzer} IRS with a single set of $T_{\rm ex}$ and $N_{\rm mol}$. Yet it is possible to obtain a reasonable fit for sub-sets of lines, that likely share similar excitation conditions. Even in the literature there is still no common agreement on the most appropriate way to fit single-slab LTE models to the water MIR emission from disks observed with \textit{Spitzer}. \citet{sal11} try to derive average gas properties for 48 T Tauri systems fitting peak-to-continuum values of many emission lines over the entire 10--35 $\mu$m range, but they also note that it is impossible to reproduce entirely the observations using a single-slab LTE model and that the outcome varies if different spectral ranges and/or fitting methods are used. \citet{cn08,cn11} instead focus on fitting lines observed in a small spectral range limited between 12 and 16 $\mu$m. Applying this method to a sample of 11 T Tauri systems \citet{cn11} report that while they can reach a very good match over this range, the same model overpredicts lines at longer wavelengths (16--19 $\mu$m). Different methods provide different results, so what is the most appropriate way to proceed and how shall we interpret different gas properties obtained fitting different spectral ranges?

A viable solution comes from theoretical works. ME09 showed that LTE models generally overpredict the line flux if H$_{2}$O is not in LTE, as emission lines excited in non-LTE regimes are sub-thermally populated \citep[see also][who explored the case of other molecules]{pavly}.  In their Figure 8, ME09 show that the difference in line flux of a water transition in the ground vibrational level excited in LTE and in non-LTE conditions is minimal for small values of $A_{ul}$. If the excitation state of the emission is unknown, a sufficient spectral resolution may allow a selection of lines to fit with LTE models based on the value of $A_{ul}$. However, they also show that such transitions are widely distributed in wavelength (Figure 7 in ME09). Therefore, when the resolution is not enough to separate different transitions (as in the case of the \textit{Spitzer} IRS) it is hard to say in advance how well a blended line can be described assuming LTE excitation. If the emission is originated in a mixture of LTE and non-LTE conditions, which is a reasonable assumption for T Tauri disks (ME09), it is therefore not surprising that LTE models may not be able to reproduce the relative strength of the observed lines. Moreover, as ME09 remark, even finding a good fit ($\chi^{2}_{red} \approx 1$) to a handful of blended water lines does not necessarily indicate that the LTE model provides a good estimate of the real conditions of the emitting gas. It is clear that some criteria for the fitting method are needed in order to use LTE models and give reliable results that do not misinterpret the data. 

A sensible choice is to use LTE models to fit only lines at the shortest wavelengths, where the overprediction of the line flux assuming LTE is minimal (Figure 7 in ME09). This is consistent with the approach utilized by \citet{cn08,cn11}. In this approach one should check the value of $E_{u}$ and $A_{ul}$ of the transitions blended in the observed lines, because transitions with high $E_{u}$ and $A_{ul}$ can be relevant even at short wavelengths (ME09). Another possibility is to consider the critical density and select lines based on the lowest values of $n_{\rm crit}$, which are the first to be thermalized if the ambient gas density is high enough. However, collision rates of water transitions are accurate only within a factor of 3--10 depending on $E_{u}$ \citep{fajos}, which in turn affects $n_{\rm crit}$ and its use as a selective tool.

\subsection{Fitting Method Adopted in This Work for H$_{2}$O Emission} \label{sec: model3}
In this work we chose to fit those H$_{2}$O lines that are \textit{the least likely to be affected by non-LTE effects}. Based on the arguments presented in Section \ref{sec: model2}, we start fitting over the short wavelengths (12--15 $\mu$m) and then we explore changes in the outcomes adding/removing lines in the range 12--19 $\mu$m. Each time a best-fit model is found for a sub-set of lines, we check the $E_{u}$, $A_{ul}$, and $n_{\rm crit}$ of the transitions that contribute for at least 10\% to the line fluxes. For $n_{\rm crit}$, we take the collision rates from \citet{fajos} and we consider collisions with H$_2$ only. We try to avoid lines that include transitions with $E_{u} \gtrsim 4300$ K or $A_{ul} \gtrsim 1$ s$^{-1}$, as in such cases non-LTE effects are more likely to be strongly relevant (ME09). To check our best-fit models we consider two factors: the value of the $\chi^{2}_{red}$ (that we require to be less than 1.5) and the plot of the ratios of observed and model line fluxes against the wavelength. More precisely, we expect the latter to resemble the behavior reported in Figure 7 in ME09 \textit{if} the fitted lines are in fact the closest to a thermal excitation\footnote{Note that we plot the ratio of observed/LTE fluxes while ME09 plot non-LTE/LTE fluxes.}. In other words, an LTE model should be able to reproduce the flux of lines given by transitions with the lowest values of $E_{u}$, $A_{ul}$, and $n_{\rm crit}$, while in the case of transitions with higher values of $E_{u}$, $A_{ul}$, and $n_{\rm crit}$ the line flux should be \textit{strictly} overpredicted. 

\begin{figure*}
\includegraphics[width=0.5\textwidth]{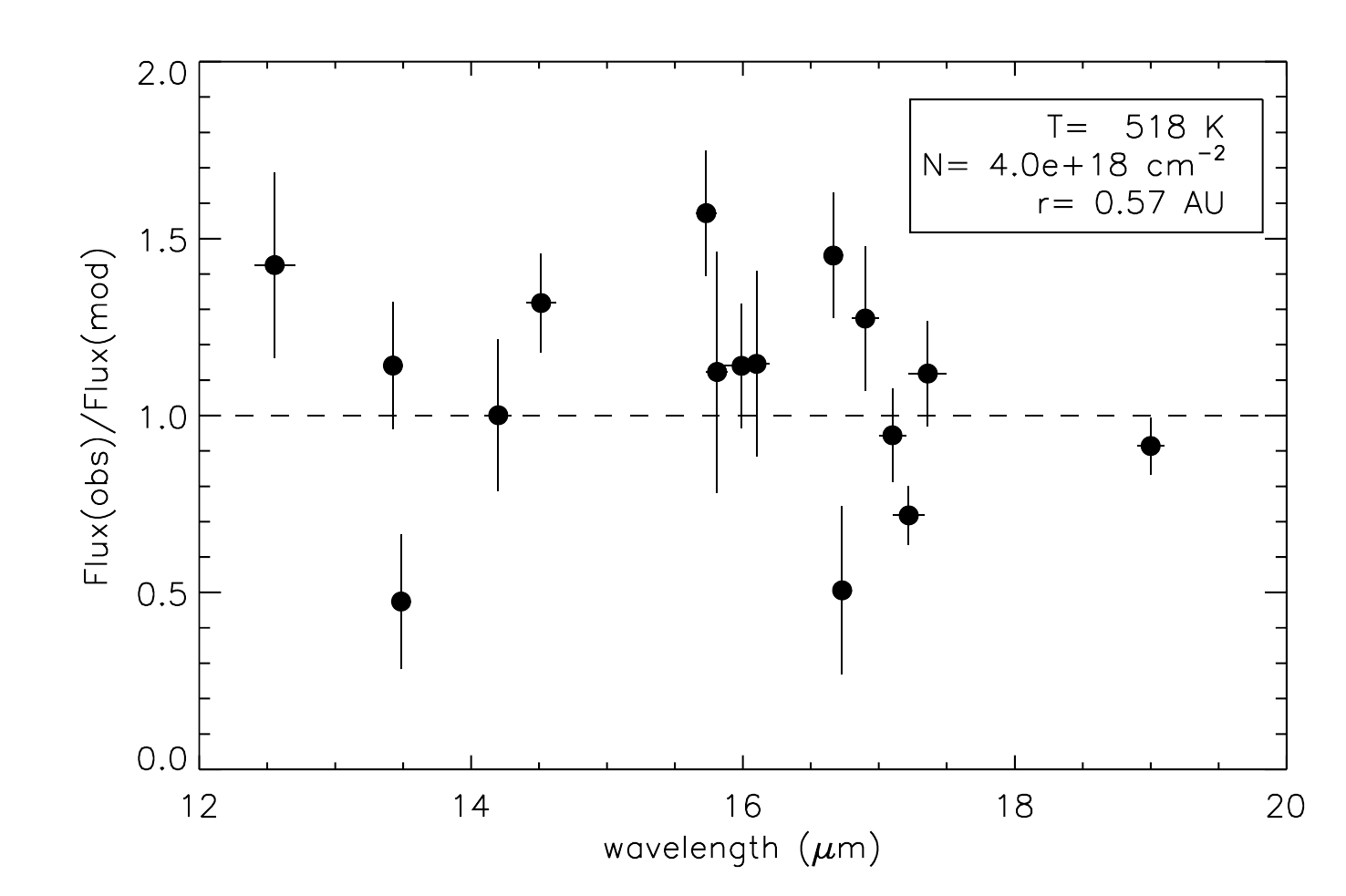} 
\includegraphics[width=0.5\textwidth]{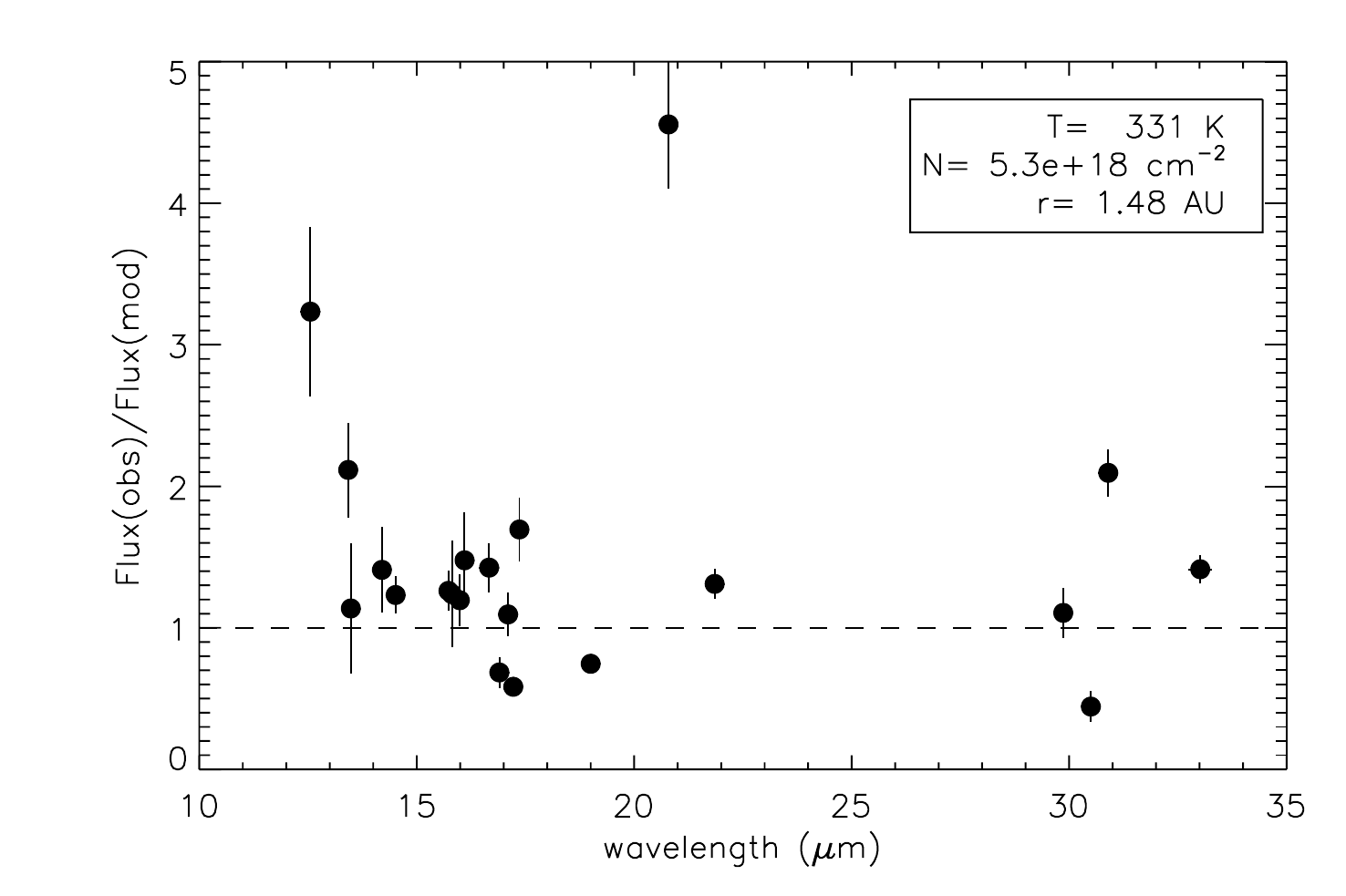} 
\caption{Quiescence: ratios of the observed and LTE model water line fluxes for the model fits to lines in SH (left) and over the entire IRS spectrum (right). The model parameters are reported in the box.}
\label{fig:res_qui}
\end{figure*}

\begin{figure*}
\includegraphics[width=0.5\textwidth]{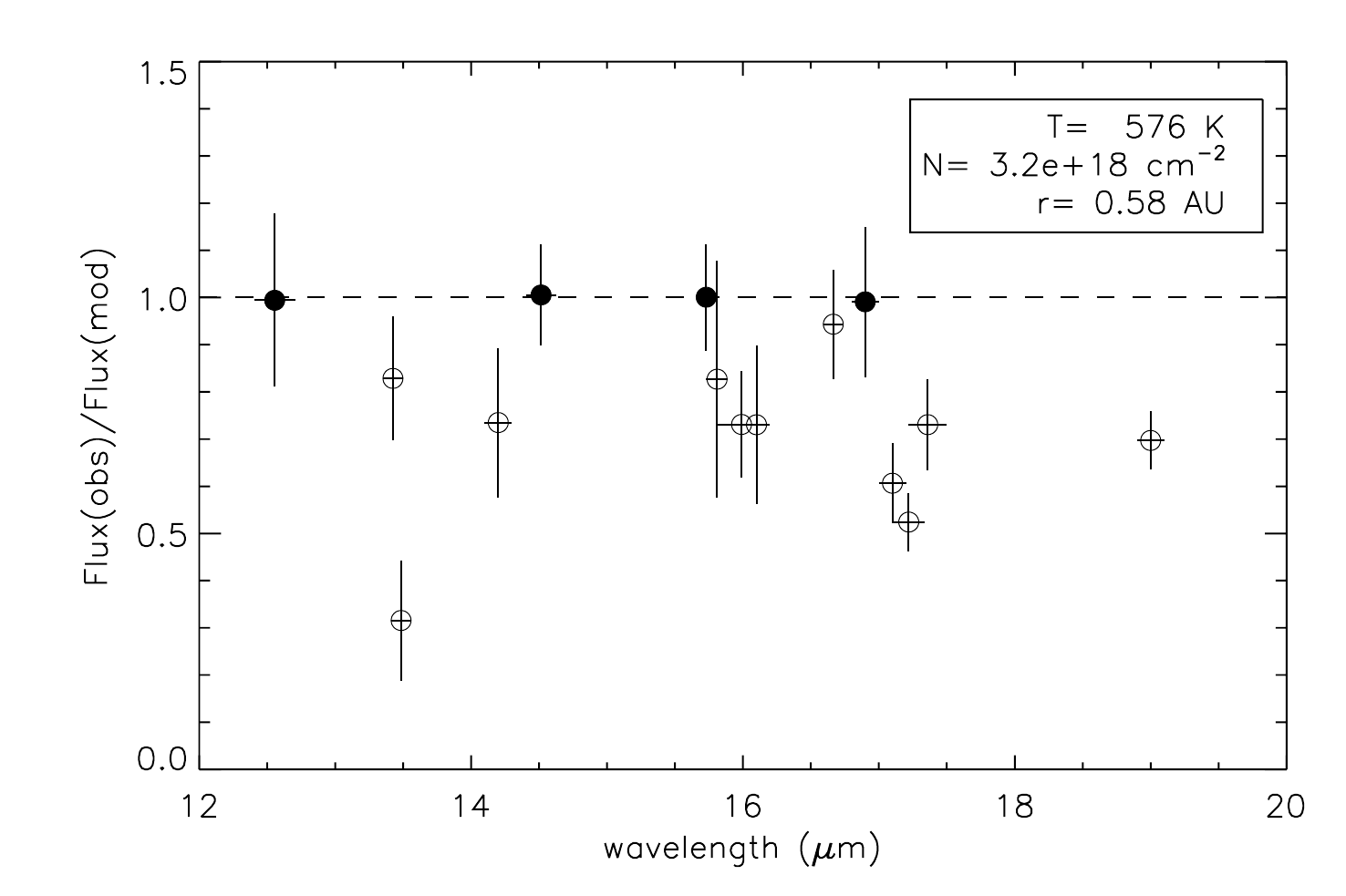} 
\includegraphics[width=0.5\textwidth]{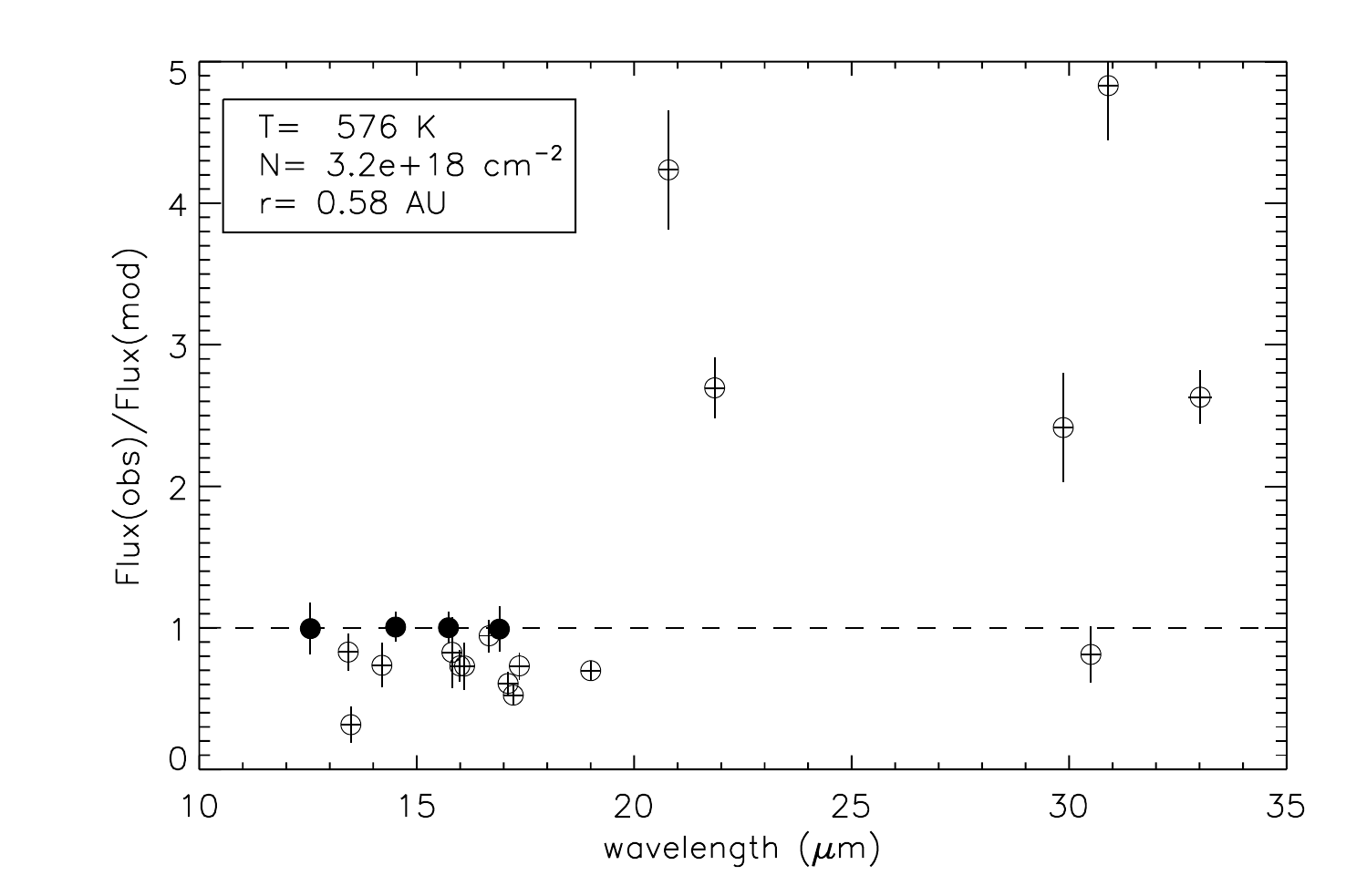} 
\caption{Same as Figure \ref{fig:res_qui}, but showing the best-fit found using only the lines indicated by filled circles, which are the least likely to be affected by non-LTE effects. All other lines, shown as empty circles, are consistent with being closer to non-LTE conditions and are overpredicted by the LTE model, while in LH the best-fit found in SH is clearly not appropriate.}
\label{fig:res_qui_2}
\end{figure*}

It is clear that our approach should be considered exploratory. Moreover, the fact that a single-slab LTE model fails to reproduce the observations may imply that the observed emission is indeed excited in non-LTE conditions \textit{and/or} that it may be produced in several disk regions with different $T_{\rm ex}$, $N_{\rm mol}$, and emitting areas rather than in a single homogeneous layer. Given the limitations of using a single-temperature LTE model, we are able to constrain only the emission that originates in a portion of the disk atmosphere where the local gas density is high enough for thermalization, and which is narrow enough to be approximated with one temperature. However, thanks to the test over the line ratios as explained in the previous paragraph, we are able to obtain indications on whether the emission is mainly to be attributed to non-LTE excitation rather than to different temperatures from different regions in the disk (see Section \ref{sec: res1}).

\begin{figure*}
\includegraphics[width=0.5\textwidth]{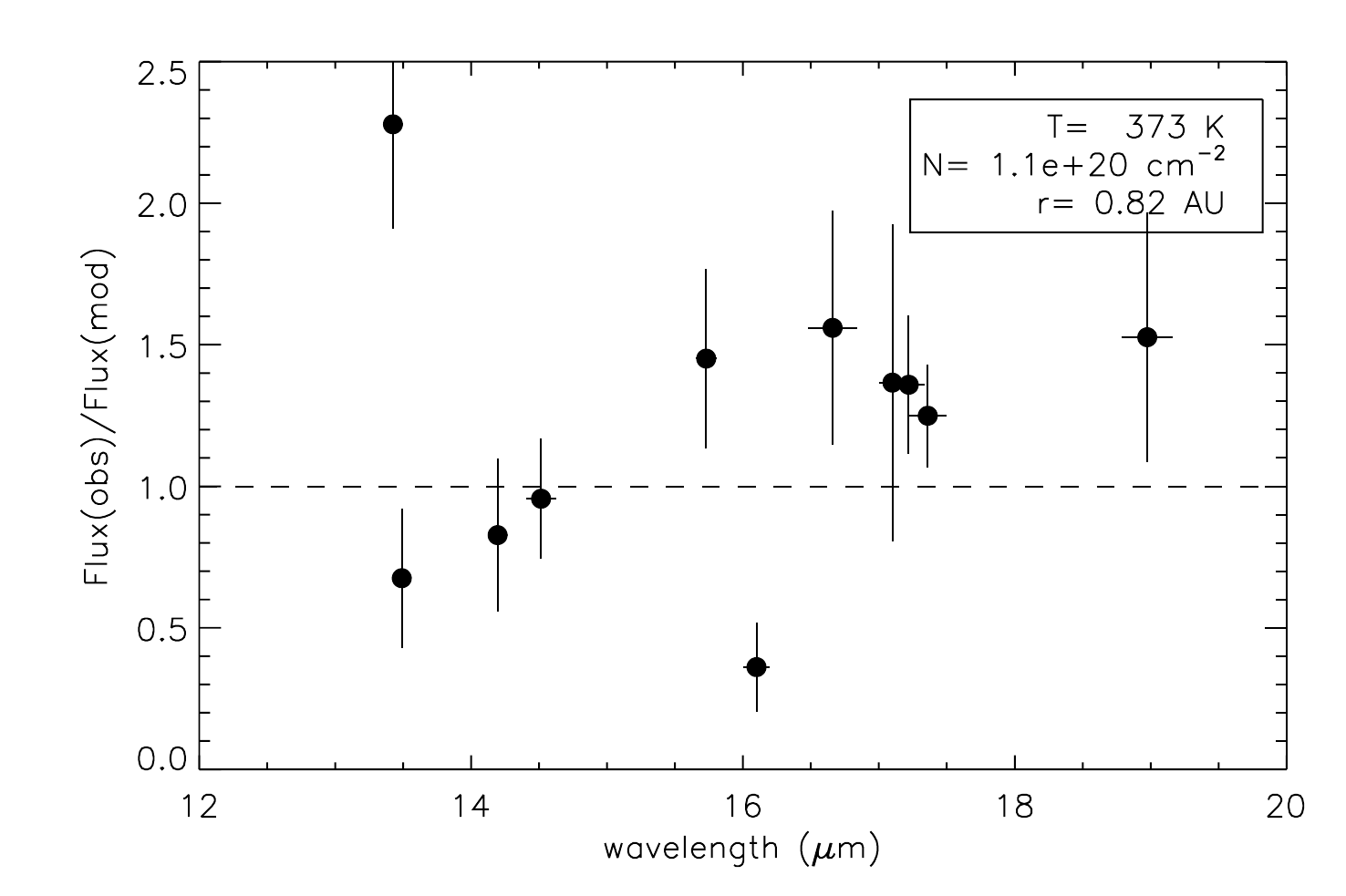} 
\includegraphics[width=0.5\textwidth]{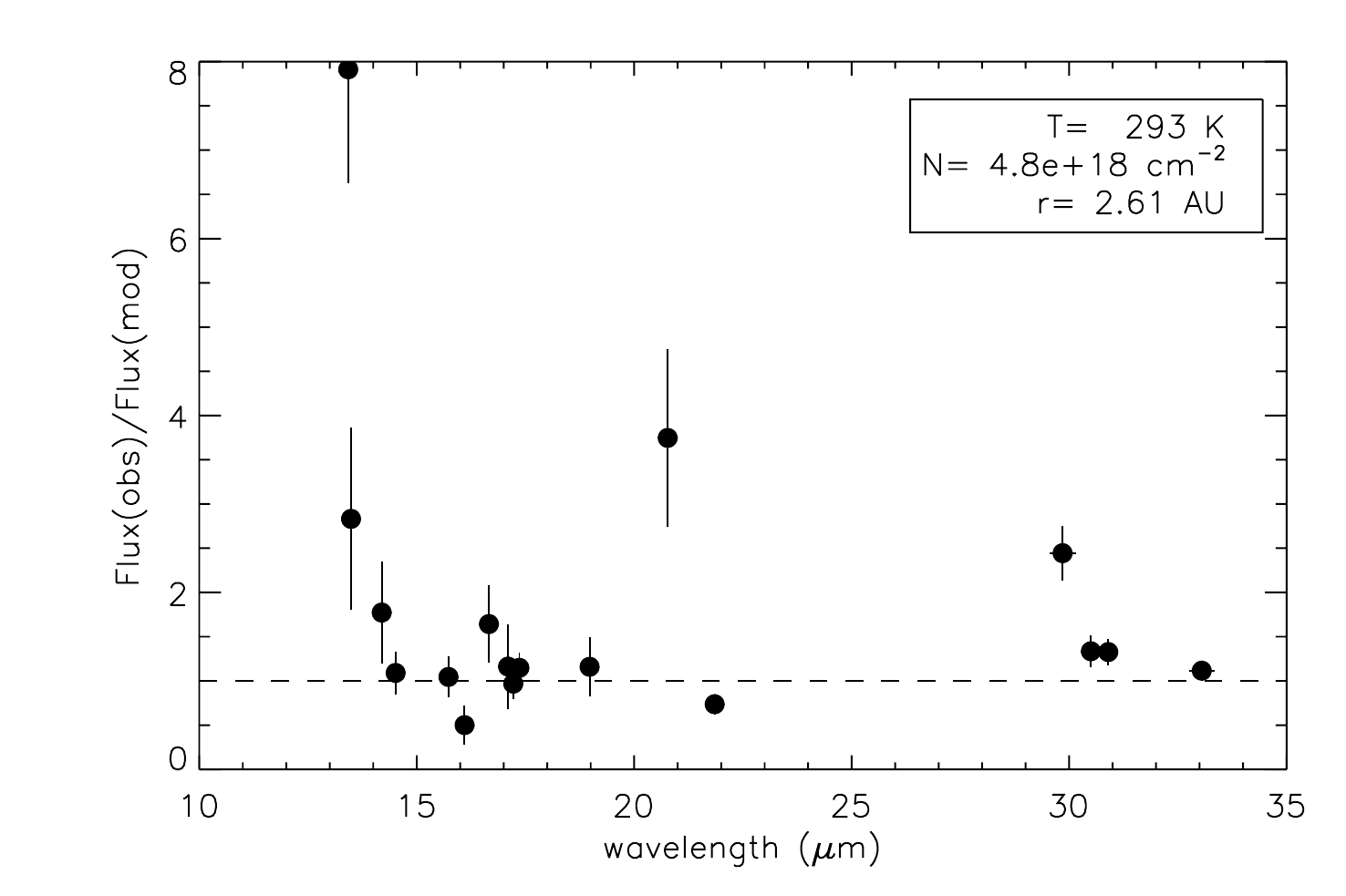} 
\caption{Outburst: ratios of the observed and LTE model water line fluxes for the model fits to lines in SH (left) and over the entire IRS spectrum (right). The model parameters are reported in the box. Water lines that are heavily contaminated by other species are not considered here.}
\label{fig:res_out}
\end{figure*}

\begin{figure*}
\includegraphics[width=0.5\textwidth]{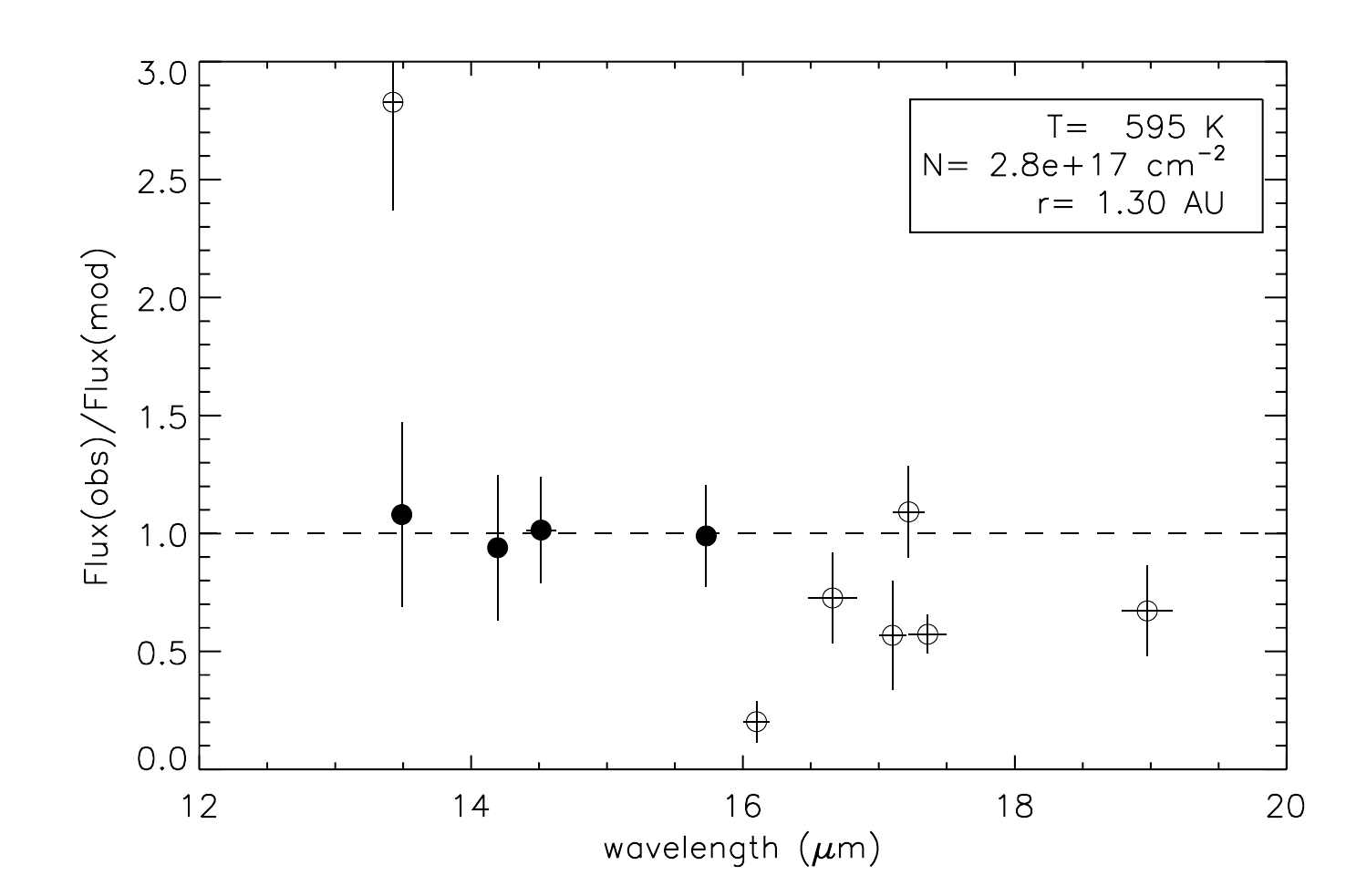} 
\includegraphics[width=0.5\textwidth]{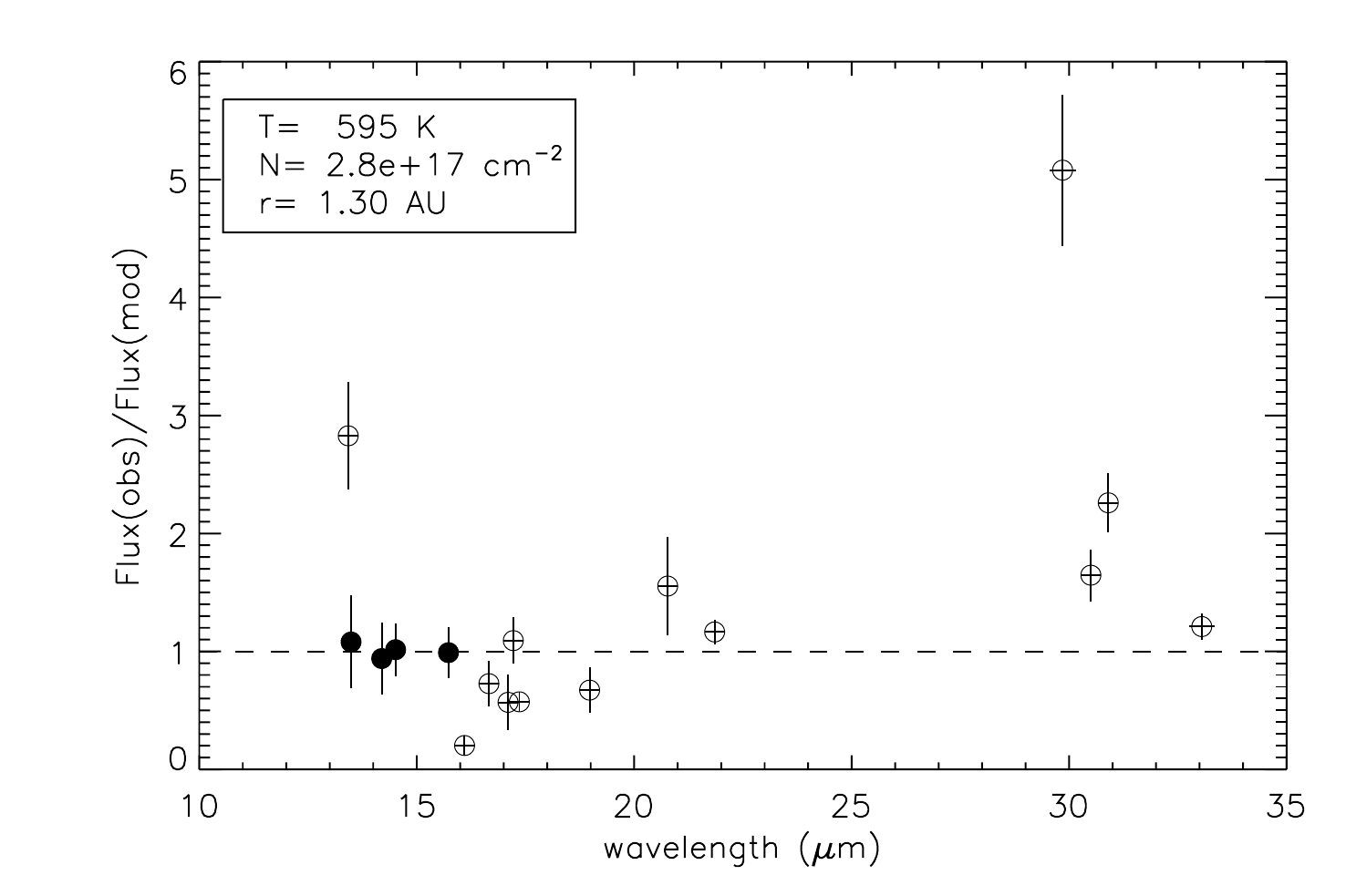} 
\caption{Same as Figure \ref{fig:res_out}, but showing the best fit found using only the lines indicated by filled circles, which are the least likely to be affected by non-LTE effects. All other lines, shown as empty circles, are consistent with being closer to non-LTE conditions and are overpredicted by the LTE model, while in LH the best-fit found in SH is clearly not appropriate. The line at $\sim13.43$ $\mu$m is strongly underpredicted by this model and is excluded from the fit.}
\label{fig:res_out_2}
\end{figure*}

\section{RESULTS OF MODELLING MOLECULAR EMISSION} \label{sec: res}
\subsection{H$_{2}$O Rotational Lines} \label{sec: res1}
\begin{figure*}
\includegraphics[width=0.33\textwidth]{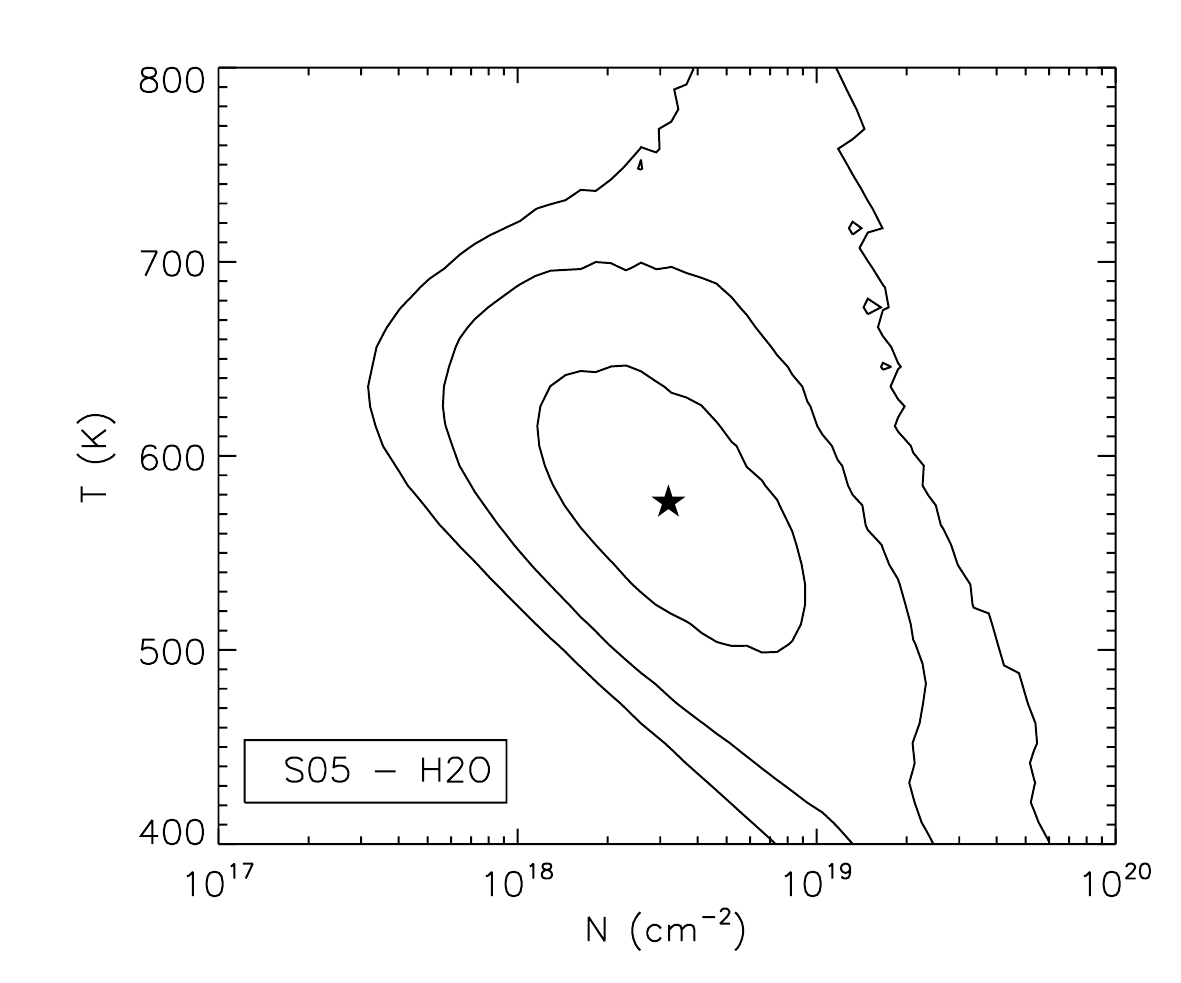} 
\includegraphics[width=0.33\textwidth]{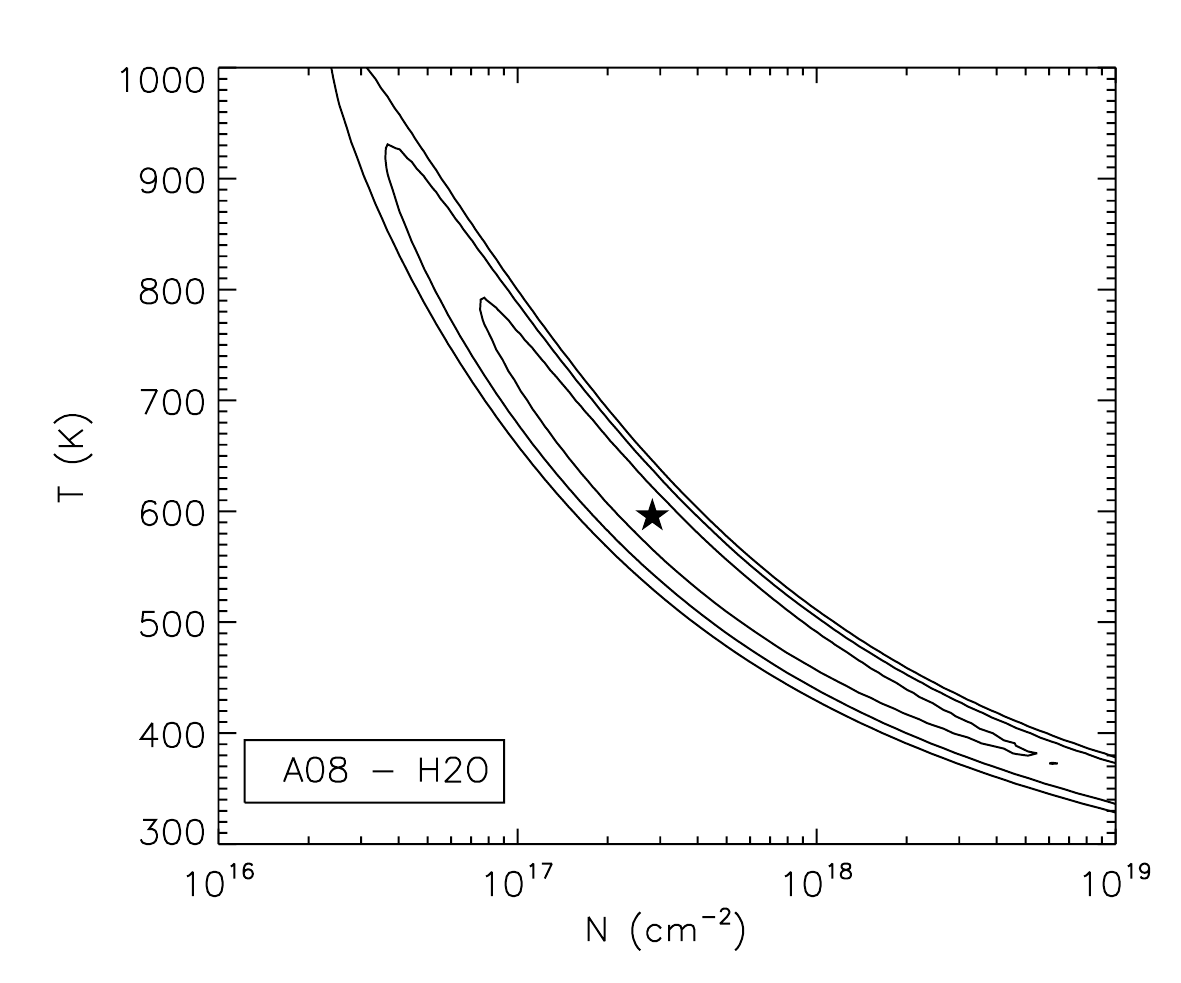} 
\includegraphics[width=0.33\textwidth]{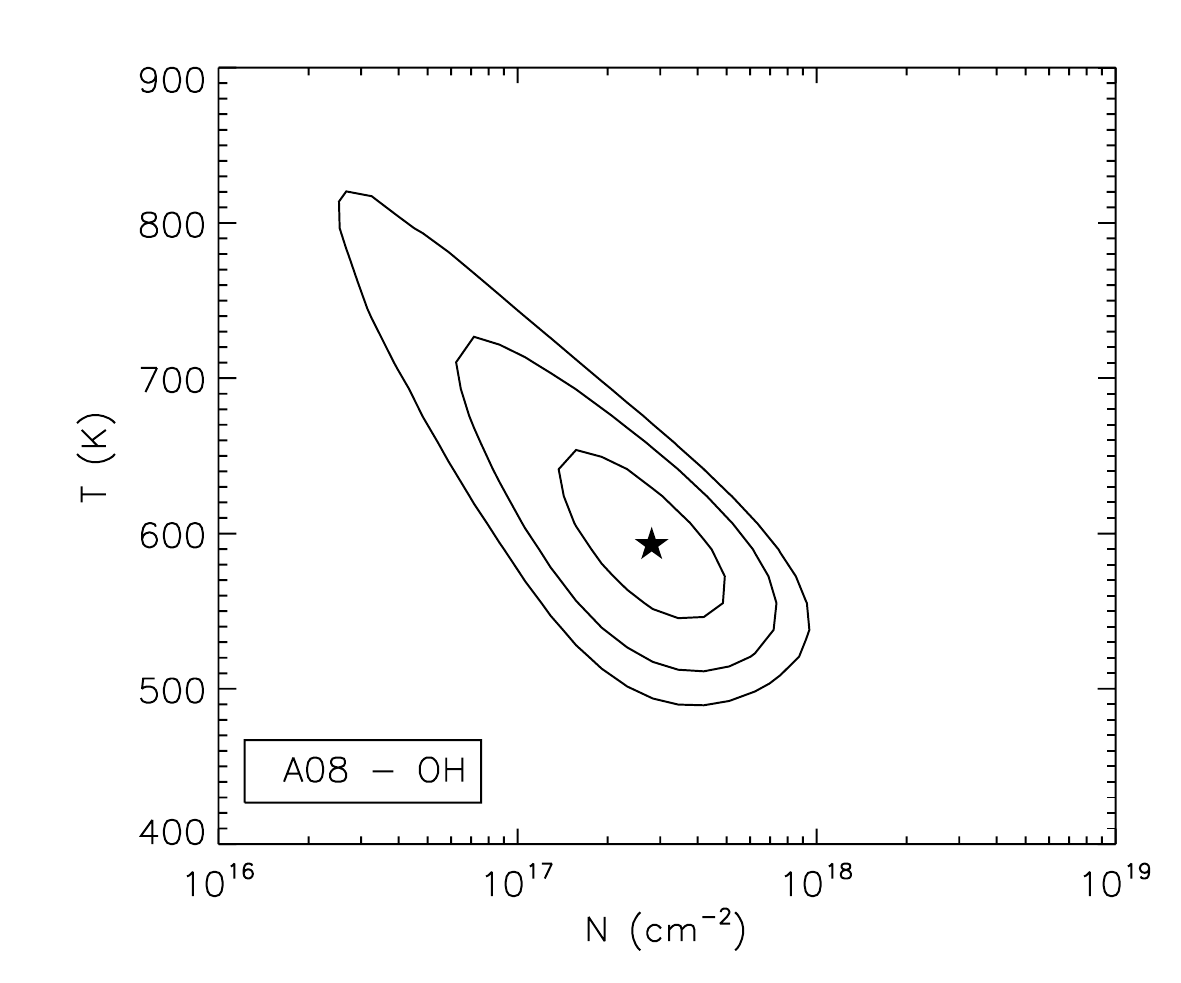} 
\caption{Confidence limits on $T_{\rm ex}$ and $N_{\rm mol}$ estimated for H$_{2}$O in quiescence (S05, left) and in outburst (A08, middle), and for the low-$E_u$ OH component in outburst (right). In the case of water in outburst we show the contours for a fixed emitting area (with $r_{p}$=1.3 AU, see Section \ref{sec: res1}). Contours show the 68\%, 95\% and 99\% two-dimensional confidence regions, while the best-fit locus is marked with a star. The degeneracy between the two model parameters is apparent, giving higher/lower temperatures for lower/higher densities.}
\label{fig:chi2}
\end{figure*}

For the sake of comparison with the results obtained following our ``LTE-fitting" method explained above in Section \ref{sec: model3}, we first report the results obtained fitting lines regardless of the values of $E_{u}$, $A_{ul}$, and $n_{\rm crit}$. In quiescence, the results obtained from fitting water lines over the SH module first, and then over the entire SH+LH coverage, are shown in Figure \ref{fig:res_qui}. The temperature varies from $\sim520$ to $\sim330$ K in the two cases respectively, while the column density stays close to $\sim 5 \times 10^{18}$ cm$^{-2}$. The probability that the data are well described by the models is lower than 0.001\% in both cases, i.e. the LTE model is strongly inappropriate as the scatter in Figure \ref{fig:res_qui} suggests. We therefore proceeded, as described above, with fitting sub-sets of lines at short wavelengths, and we found a best fit over the range 12--17 $\mu$m with $T_{\rm ex} = 580$ K and $N_{\rm mol} = 3.2 \times 10^{18}$ cm$^{-2}$ (see Figure \ref{fig:res_qui_2}). The $\chi^{2}_{red}$ is much lower than 1, suggesting that our error-bars may overestimate the real errors by a factor of $\sim$ 5 or more. The lines used in the best fit are predominantly produced by transitions with $E_{u} \lesssim 4300$ K and $A_{ul} \lesssim 1$ s$^{-1}$, and their critical densities range between $10^{8}$ and a few $10^{9}$ cm$^{-3}$ (see Table \ref{tab:water_fits_par}). Therefore the chosen lines are the first to thermalize, compared to other lines in the SH module that instead include transitions with $E_{u}$ up to 7000 K, $A_{ul}$ up to $10^{2}$ s$^{-1}$, and $n_{\rm crit}$ up to $10^{11}$ cm$^{-3}$. The behavior of the line flux ratios plotted in Figure \ref{fig:res_qui_2} supports that we are indeed probing the most thermalized lines: while the LTE model can reproduce very well the fitted lines, other lines are instead \textit{strictly} overpredicted.  The emission is close enough to an optically thick regime ($\tau \sim$ 1--10) that the emitting area is well constrained and the best fit is found for $r_{\rm p}=$ 0.58 AU. The confidence limits derived for the best-fit parameters are shown in Figure \ref{fig:chi2} and listed in Table \ref{tab:conf_lim}.

\begin{deluxetable*}{l l c c c c}
\tabletypesize{\small}
\tablewidth{500pt}
\tablecaption{\label{tab:water_fits_par}Parameters of H$_{2}$O Lines Used to Constrain the Emission Observed in EX Lupi.}
\tablehead{\colhead{Line ($\mu$m)} & \colhead{Transitions ($\mu$m)} & \colhead{$E_{u}$ (K)} & \colhead{$A_{ul}$ (s$^{-1}$)} & \colhead{$n_{crit}$ (cm$^{-3}$)} & \colhead{$\tau$} }
\startdata
\cutinhead{Quiescence (S05)} 
12.51 & 12.51 & 4133 & 1.50 & 4 (9) & 0.6 \\
         & 12.52 & 4133 & 1.50 & 3 (9) & 1.7 \\
14.51 & 14.49 & 3234 & 1.22 & 3 (9) & 2.1 \\
         & 14.51 & 3233 & 1.21 & 2 (9) & 6.2 \\
         & 14.54 & 2605 & 0.15 & 3 (8) & 1.6 \\
15.74 & 15.74 & 2824 & 1.09 & 2 (9) & 11.4 \\
16.89 & 16.89 & 2698 & 0.70 & 2 (9) & 8.2 \\
         & 16.90 & 2125 & 0.26 & 5 (8) & 7.5 \\

\cutinhead{Outburst (A08)} 
13.49 & 13.48 & 5499 & 8.61 & 2 (10) & 0.1 \\
         & 13.50 & 3341 & 0.49 & 1 (9) & 0.2 \\
14.20 & 14.18 & 2876 & 0.30 & 6 (8) & 0.2 \\ 
         & 14.21 & 3951 & 3.35 & 8 (9) & 0.5 \\
14.51 & 14.49 & 3234 & 1.22 & 3 (9) & 0.2 \\
         & 14.51 & 3233 & 1.21 & 2 (9) & 0.6 \\
         & 14.54 & 2605 & 0.15 & 3 (8) & 0.2 \\
15.74 & 15.74 & 2824 & 1.09 & 2 (9) & 1.1 
\enddata
\tablecomments{For each observed line, we report the transitions that, according to the best-fit model, contribute for more than 10\% to the observed flux. The value of the opacity of each transition is dependent on the best-fit model (in outburst we report here the model with $r_{\rm p}=1.3$ AU). All other parameters are instead intrinsic of each single transition. Molecular data are taken from the HITRAN database, apart from the collision rates which are taken from \citet{fajos}. Critical densities must be multiplied by $10^{a}$, where the exponent $a$ is shown in parentheses.}
\end{deluxetable*}

The same procedure is applied to the outburst spectrum, and the results are shown in Figures \ref{fig:res_out} and \ref{fig:res_out_2}. Here we have an additional limitation given by the increased confusion of water lines with other molecules/atoms, mainly OH and H \textsc{i} at short wavelengths (see Figure \ref{fig:spec_comp}). Therefore, from the sample of lines that were used in quiescence we exclude those which are the most contaminated by other species (see Table \ref{tab:fluxes}). Fitting over the SH module gives a temperature of $\sim$ 370 K and a column density of $\sim 10^{20}$ cm$^{-2}$, while over the entire spectrum we find a temperature of $\sim$ 300 K and a column density of $\sim 5 \times 10^{18}$ cm$^{-2}$. Again, the model is not appropriate to reproduce the observations (probability lower than 0.001\%), as the scatter in Figure \ref{fig:res_out} shows. The best-fit model using instead only lines closer to thermalization gives a good fit with $T_{\rm ex} \approx 600$ K and $N_{\rm mol} \approx 3 \times 10^{17}$ cm$^{-2}$, but the emission is more optically thin than in outburst ($\tau \sim$ 0.1--1) and it is sensitive to changes in the emitting area only for low values of $r_{\rm p}$ (i.e. high values of $N_{\rm mol}$, when the emission gets optically thick). The emission is found consistent with $r_{\rm p}\gtrsim 0.6$ AU (68\% confidence), but for $r_{\rm p} > 2$ AU it is so optically thin ($\tau <$ 0.1) that the result is completely insensitive to a change in emitting area. Assuming that OH and H$_{2}$O emission probes the same disk radii \citep[as recently shown from very high-resolution near-infrared lines by][]{dopp}, we choose to fix the emitting area for H$_{2}$O in outburst to $r_{\rm p}=1.3$ AU (the best-fit value of low-$E_{u}$ OH in outburst, see Section \ref{sec: res2}). Then we derive confidence limits on $T_{\rm ex}$ and $N_{\rm mol}$ for this fixed radius, which we take as our reference case (see Table \ref{tab:conf_lim} and Figure \ref{fig:chi2}). The opportunity of having the same emitting area for OH and H$_{2}$O is that we can easily compare their relative change in terms of $T_{\rm ex}$ and $N_{\rm mol}$ only. In the case that the assumption was found to be wrong and H$_{2}$O had instead the same radius found in quiescence (0.58 AU), the model gives $T_{\rm ex} \approx 700$ K and $N_{\rm mol} \approx 1 \times 10^{18}$ cm$^{-2}$ to explain the increase in line fluxes. 
In our reference case, fixing $r_{\rm p}$ to 1.3 AU, the lines considered in the best fit have mostly transitions with ranges of $E_{u}$, $A_{ul}$, and $n_{\rm crit}$ consistent with the quiescent case, even though the best-fit model shows a relevant contribution from transitions having higher values of $A_{ul}$, and $n_{\rm crit}$ (see Table \ref{tab:water_fits_par}). The behaviour of line ratios discussed above is once again confirmed\footnote{A noticeable exception is the line at $\sim13.43$ $\mu$m, which, despite being one with the lowest $E_{u}$ and $A_{ul}$, is underpredicted by the best-fit model. Contamination by another species is a viable explanation.} (Figure \ref{fig:res_out_2}). However, it is clear that the sub-set of lines we end up fitting in outburst is more affected by non-LTE excitation, and that the $T_{\rm ex}$ derived is therefore probably only a lower limit to the real gas temperature of the observed emission.

In fact, $T_{\rm ex}$ always lowers when we include in the fits lines that may be more affected by non-LTE according to the criteria explained in Sections \ref{sec: model2} and \ref{sec: model3}. It is very interesting that a similar effect is seen in comparing two recent studies. Despite the similarity in the samples of T Tauri stars considered in the two papers, the average gas temperature reported in \citet{sal11}, who include in their fits lines at wavelengths up to 35 $\mu$m, is $\approx$ 25\% lower than what \citet{cn11} find, who instead limit their fits to the 12--16 $\mu$m range. A viable explanation is that the different results are due to a different contribution from non-LTE excitation given by the different choice of the spectral ranges to fit. This is supported by the fact that, using the outburst spectrum, our result obtained from fitting lines over the entire spectrum is consistent with what \citet{sal11} found for EX Lupi in outburst, while when we restrict the fits to lines closer to LTE the results are consistent with the average found by \citet{cn11}. As already said, if transitions at long wavelengths suffer more from non-LTE effects (i.e. are sub-thermally populated), then the $T_{\rm ex}$ found using an LTE model fitted to such transitions will be lower than if they were properly modelled in non-LTE. The fact that a lower temperature is always found when lines at long wavelengths are included strongly suggests that the MIR emission we observe is indeed from non-LTE excitation, and that only some lines can be approximated as being in LTE.

However, non-LTE excitation is not alone in affecting MIR molecular emission. In fact, the best fits we obtain from fitting lines closer to LTE are clearly not appropriate for the emission lines detected in LH: line fluxes are strongly underpredicted, contrarily to what is found in the SH module (see Figures \ref{fig:res_qui_2} and \ref{fig:res_out_2}). Lines at wavelengths longer than 20 $\mu$m are predominantly produced by lower $E_{u}$ than at shorter wavelengths, typically $E_{u}\sim$ 900-1800 K. Therefore an LTE solution with $T_{\rm ex}\sim$ 200-300 K could in principle account for such low levels. However, the strict overprediction of line fluxes that we find using lines in the SH module cannot be found using lines in LH \textit{even} if we assume a low temperature. A reason for that can be seen in that the $A_{ul}$ are instead generally higher ($A_{ul}\gtrsim10$ s$^{-1}$) than at short wavelengths, and higher is in turn the effect of non-LTE excitation. This is again consistent with Figure 7 in ME09: at wavelengths longer than 20 $\mu$m the ratios of non-LTE and LTE fluxes span the whole range from 1 to 0, such that no line observed with the \textit{Spitzer} resolution can be approximated as in LTE. Nonetheless, non-LTE excitation alone is not able to account for the observed fluxes at long wavelengths when we assume the same temperature found at the short wavelengths. Given the lower $E_{u}$, what we observe in the LH module is likely produced by water vapor emitted from a colder region in the disk, probably at larger radii.

\subsection{OH Rotational Lines} \label{sec: res2}
\begin{figure}
\includegraphics[width=0.47\textwidth]{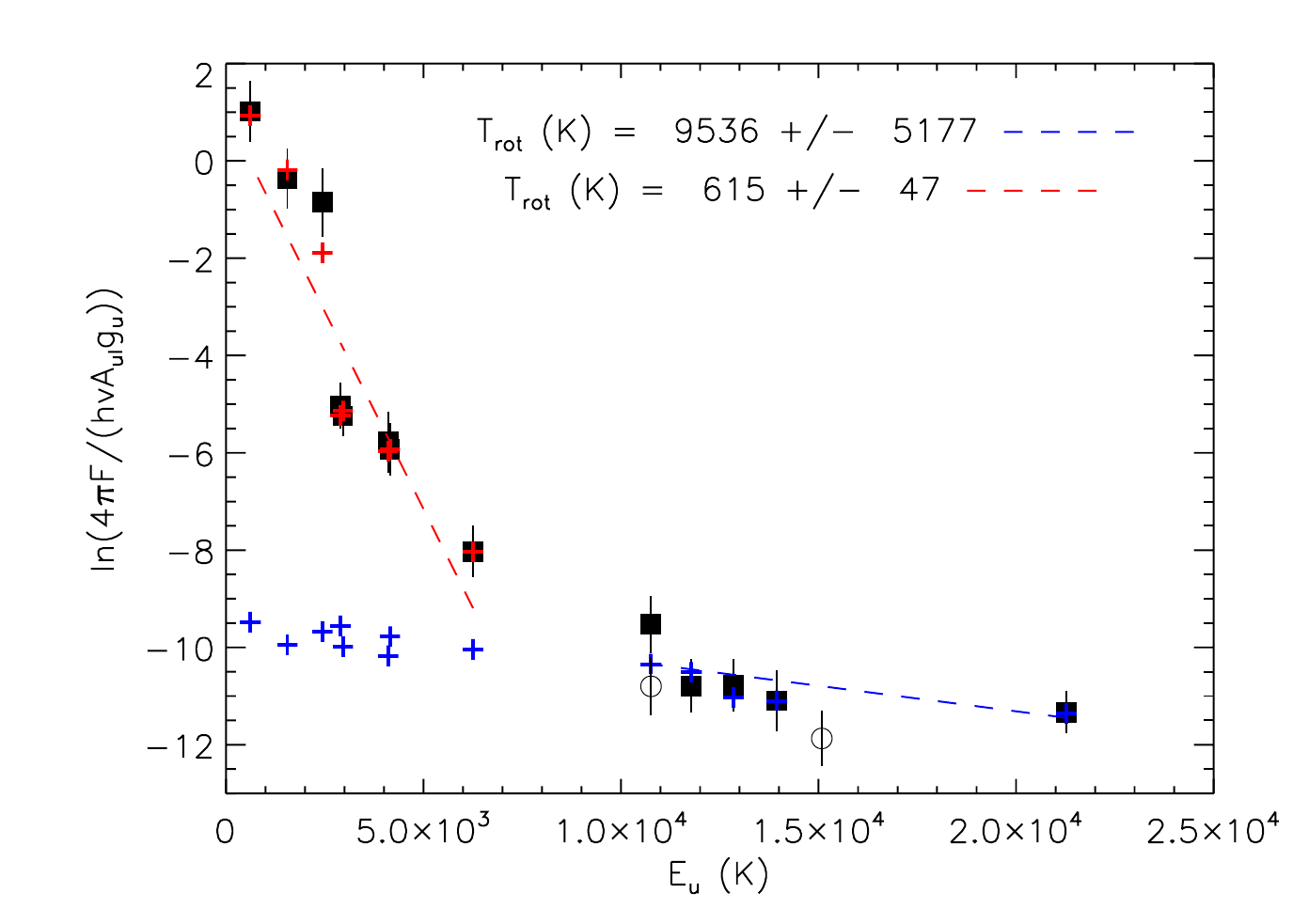} 
\caption{Rotation diagram of OH lines detected in outburst (see Table \ref{tab:oh}), reported as black squares with error-bars. Linear fits suggest two separate emission components (see Section \ref{sec: res2}), and are shown with red/blue dotted lines. Two LTE models can reproduce the line fluxes of each component separately (red/blue crosses), but neither one is able to reproduce all lines together (red crosses fall outside the plot in the high-$E_u$ range). Consideration of the two detections in quiescence (empty circles with error-bars) suggests a lower column density and a temperature consistent with outburst for the component given by high-$E_u$ lines.}
\label{fig:rot_diagr}
\end{figure}

OH rotational transitions from different levels are not as intermingled in wavelength as for the H$_{2}$O molecule, and the individual lines observed with \textit{Spitzer} are composed of doublets from a very narrow range of $E_{u}$ and $A_{ul}$ (see Table \ref{tab:oh}). This enabled us to perform a rotation diagram analysis, using the description in \citet{larss}. In Figure \ref{fig:rot_diagr} we show the rotation diagram based on the OH lines detected in outburst. We assume that their measured fluxes are entirely given by OH emission, and we exclude the $\sim$16.66 $\mu$m doublet which is strongly blended with water (see Figure \ref{fig:cl_trans}). It is apparent from the rotation diagram that the detected lines are not well approximated with a single straight line, that would be expected if the emission was optically thin and in LTE. GL99 showed how deviations can be due to different factors: LTE/non-LTE excitation, optically thin/thick emission, the complexity of the molecular structure, and a multi-temperature emission. The interpretation of the emission we observe in EX Lupi is not straightforward because we may have a combination of effects. First of all, the OH molecule does not belong to the simple linear molecules studied in GL99, given its ground-state double-ladder excitation structure. Second, we very likely observe emission that can only partially be approximated to LTE, as in the case of H$_{2}$O. Third, the opacity of emission lines is not known \textit{a priori} and depends on $T_{\rm ex}$ and $N_{\rm mol}$. Unambiguously distinguishing between these different effects is challenging. We propose here a viable interpretation. 

Given the much smaller transition probabilities of the cross-ladder compared to the intra-ladder transitions (see Table \ref{tab:oh}), the former are the most likely to be optically thin. A linear fit to the cross-ladder fluxes alone finds $T_{\rm rot} \approx 950 \pm 460$ K, but we note that a reasonable fit could also include lines up to $E_{u}\sim6000$ K. If we extend the linear fit to include such lines we find $T_{\rm rot} \approx 620 \pm 50$ K (see Figure \ref{fig:rot_diagr}). It is instead apparent from the diagram that such a low rotational temperature is certainly not adequate for higher-$E_{u}$ lines. When fitting separately over the high-$E_{u}$ range ($E_{u}>10,000$ K) we find $T_{\rm rot} \approx 9500 \pm 5200$ K (see Figure \ref{fig:rot_diagr}). Two components can explain the observed line fluxes even when we account for the optical depth using our LTE model presented in Section \ref{sec: model1}. A model fit to lines with $E_{u} \lesssim 6000$ K finds $T_{\rm ex} = 590$ K and $N_{\rm mol} = 2.8 \times 10^{17}$ cm$^{-2}$ over an emitting area with $r_{\rm p}=$ 1.30 AU. The emission is optically thick (although the cross-ladder transitions are still optically thin) and well characterized (see Figure \ref{fig:chi2} and Table \ref{tab:conf_lim}), so that we take its area to be the same for water in outburst as assumed in Section \ref{sec: res1}. On the other hand, also the high-$E_{u}$ lines ($E_{u} > 10,000$ K) can be reproduced reasonably well by a single-temperature LTE model with $T_{\rm ex} \approx 10,000$ K and $N_{\rm mol} \approx 1 \times 10^{14}$ cm$^{-2}$ (fixing $r_{\rm p}$ to 1.3 AU), the emission is very optically thin and cannot reproduce line fluxes with $E_{u} \lesssim 6000$ K. Despite the success of these two separate LTE solutions for different ranges of $E_{u}$ (see Figure \ref{fig:rot_diagr}) we do not conclude that the emission is in LTE. A few transitions lying on a straigth line can also be mimicked by a quasi-thermal behavior (GL99), and finding different temperatures when different ranges of $E_{u}$ are fitted suggests non-LTE excitation. In conclusion, this analysis shows that the emission in EX Lupi in outburst \textit{can} be explained by two LTE components, one \textit{warm} ($T\sim600$ K) and one \textit{hot} ($T\gtrsim4000$ K). It is nonetheless likely that a two-temperature LTE description is simply an approximation, and that many factors play a role in producing the scatter/curvature that we see in the rotation diagram: optical depth at low $E_{u}$, non-thermal excitation at high $E_{u}$ (in fact for $T_{\rm gas} \gtrsim 10,000$ K thermal dissociation would become important), and/or a range of gas temperatures instead of only two. 

We briefly analyze also the quiescent case. The high-$E_{u}$ component, with only two detected lines, suggests a comparable temperature and a lower column density as compared to outburst (see Figure \ref{fig:rot_diagr}). As mentioned above, the low-$E_{u}$ component is instead not detected. If we assume that the flux from the two dubious detections at $\sim 23.22$ and 27.67 $\mu$m (Figure \ref{fig:spec_comp} and Table \ref{tab:oh}) is to be attributed to OH alone and we fit these lines, we find a low-$E_{u}$ OH column density of $\approx 0.7 \times 10^{17}$ cm$^{-2}$ (assuming the same $T_{\rm ex}$ found in outburst, 590 K, and the same emitting area found for water in quiescence, with $r_{\rm p}=$ 0.58 AU).

\begin{figure}
\includegraphics[width=0.48\textwidth]{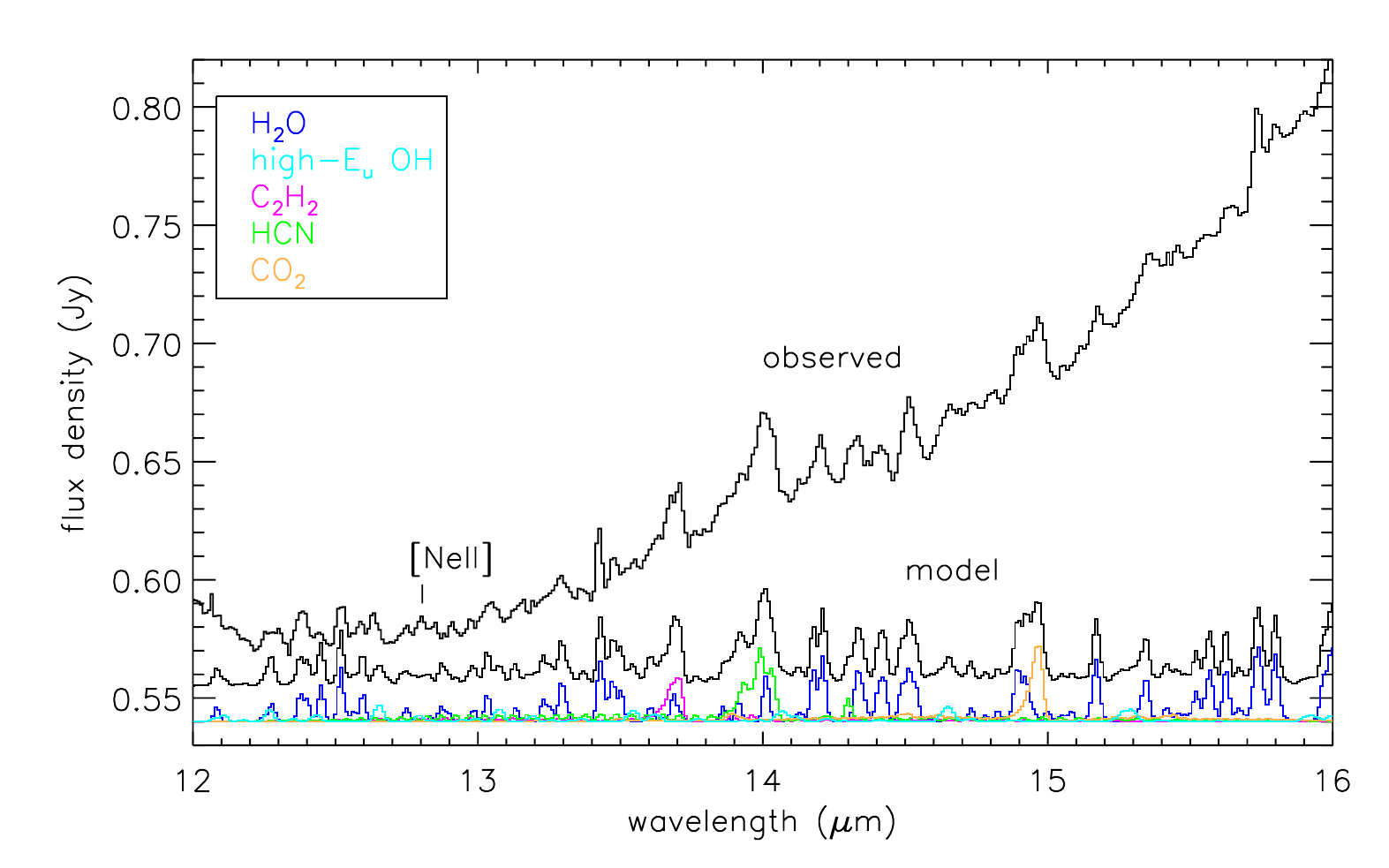} 
\caption{Quiescent spectrum (S05) and LTE models derived for H$_{2}$O (blue) and OH (cyan), plotted together with the models assumed for HCN (green), C$_{2}$H$_{2}$ (purple) and CO$_{2}$ (orange). The sum of all model contributions from different molecules is shown in black and shifted for comparison. A continuum is not added to the model. For the three organic molecules we assume the same emitting area found for water emission ($r_{\rm p}=$ 0.58 AU), and model parameters consistent with the range of values derived for other T Tauri systems by \citet{cn08,cn11}, namely $T_{\rm ex} = 750$ K and $N_{\rm mol} \sim$ 2--7 $\times 10^{15}$ cm$^{-2}$.}
\label{fig:models_qui}
\end{figure}

\subsection{C$_{2}$H$_{2}$, HCN and CO$_{2}$ Rovibrational Branches} \label{sec: res3}
The Q-branches of C$_{2}$H$_{2}$, HCN and CO$_{2}$ produce broad blended features in the \textit{Spitzer} spectra, contribute to each other's flux, and are also contaminated by H$_{2}$O and OH lines. The shape of the Q-branch, as it is seen with the \textit{Spitzer} spectral resolution, depends on both the temperature and the column density of the emitting gas. Therefore, our fitting method is not sufficient to constrain the emission from organics, as the line flux alone is very highly degenerate in $T_{\rm ex}$ and $N_{\rm mol}$. A careful analysis of the unresolved emission from these molecules requires at least that all other chemical species are well constrained and subtracted from the data, and that the Q-branch profile is accounted for in the model fitting. Two attempts in this direction have been recently shown by \citet{cn11} and \citet{sal11}. The constraints on $T_{\rm ex}$ and $N_{\rm mol}$ they are able to derive are loose such that a characterization of the change in organic emission in the case of EX Lupi between quiescence and outburst would be very hard and uncertain. Therefore in this paper we refrain from any kind of sophisticated analysis of the organic emission, and simply provide a rough estimate of the change observed between quiescence and outburst in terms of a variation in column density. 

\begin{figure}
\includegraphics[width=0.48\textwidth]{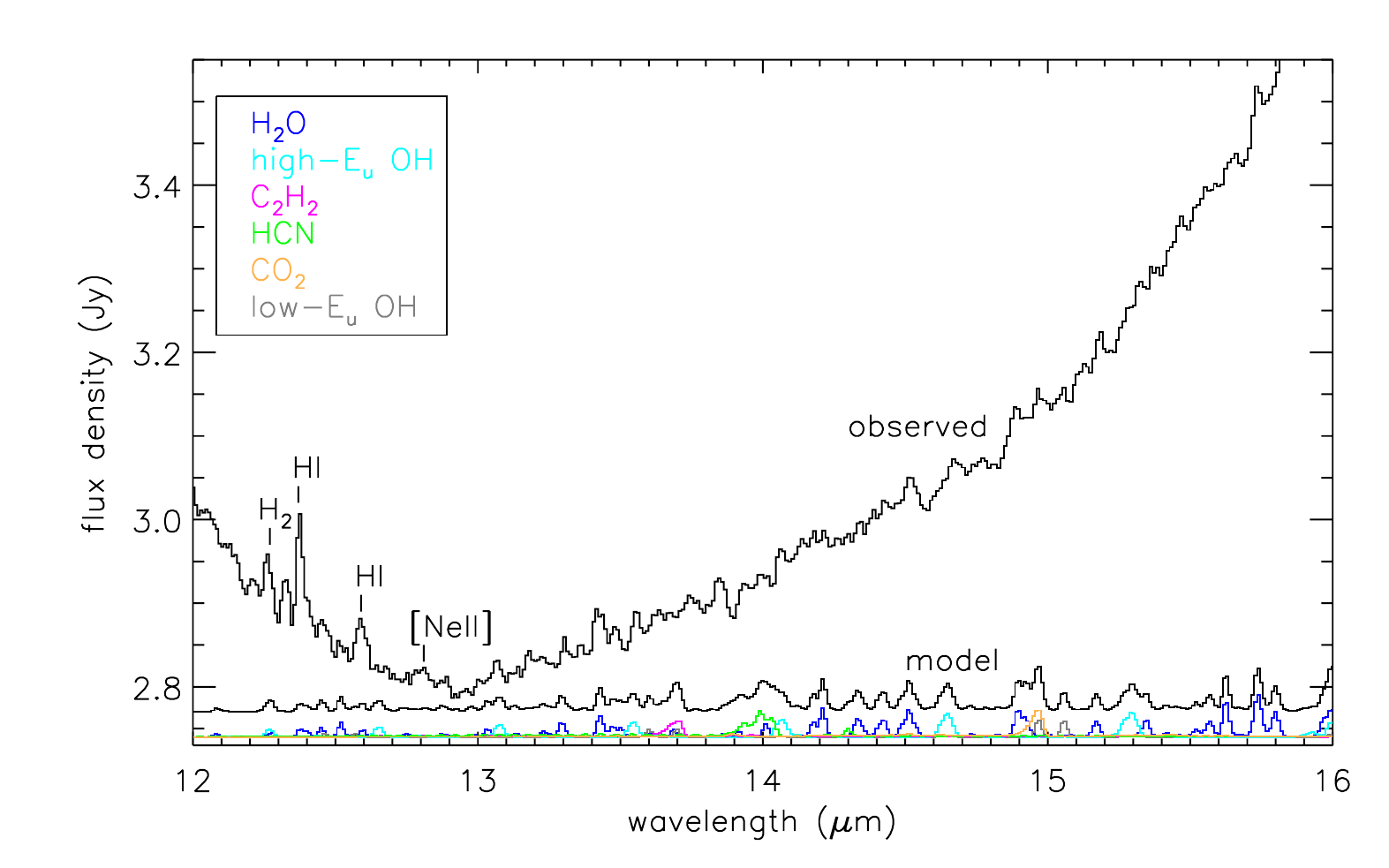} 
\caption{Same as Figure \ref{fig:models_qui} but showing the outburst (A08) spectrum. The LTE model derived for low-$E_u$ OH emission is added in grey (two lines can be seen at $\sim$15 $\mu$m). Models for organics are plotted using the same parameters assumed in quiescence. It can be seen that in outburst the Q-branches of C$_{2}$H$_{2}$ at $\sim$13.7 $\mu$m, HCN at $\sim$14 $\mu$m, and CO$_{2}$ at $\sim$15 $\mu$m are not as strong as in quiescence. Remarkable emission lines from other identified molecules/atoms are labelled. Unidentified species may be responsible for other features, like those at $\sim13.8$ $\mu$m.}
\label{fig:models_out}
\end{figure}

In Figure \ref{fig:models_qui} we show the quiescent S05 spectrum and LTE models for C$_{2}$H$_{2}$, HCN, and CO$_{2}$ \textit{assuming} parameters in the ranges derived for other T Tauri disks by \citet{cn08,cn11}. Namely, for EX Lupi we assume $T_{\rm ex} = 750$ K and $N_{\rm mol} \sim$ 2--7 $\times 10^{15}$ cm$^{-2}$, while we take the emitting area from water emission ($r_{\rm p}=$ 0.58 AU). It can be seen from the figure that the emission observed in EX Lupi does not show any remarkable difference from the models. In Figure \ref{fig:models_out} it can be seen that the models assumed for organics in quiescence are instead not consistent with the outburst A08 spectrum. The emission from C$_{2}$H$_{2}$ at 13.7 $\mu$m and HCN at 14 $\mu$m is clearly no longer as strong as it was in quiescence. The case of CO$_{2}$ is more hard to judge. The Q-branch of CO$_{2}$ at 14.97 $\mu$m is close to a strong water line at $\sim$14.9 $\mu$m, and overlaps with one OH cross-ladder doublet (see Figure \ref{fig:models_out} and Table \ref{tab:oh}). It is therefore hard to say if a small contribution from CO$_{2}$ is still present in outburst, but the observations can be explained by H$_{2}$O+OH emission alone. This is consistent with the non-detection of CO$_{2}$ (and C$_{2}$H$_{2}$ and HCN as well) reported in \citet{pont10} for EX Lupi in outburst, using the C08 spectrum. 

We then use our LTE model to add the organic spectral features in the outburst spectrum and derive upper limits. We start with the model parameters assumed in quiescence and decrease $N_{\rm mol}$ until the line fluxes get below the noise limit in outburst. The result is that all organic molecules have to be reduced in column density by at least a factor $\sim$ 5 to be undetected in outburst. If the emitting area and/or the temperature increase in outburst then this factor is $\sim$ 10 or more.
\\

\begin{deluxetable*}{c l l l}
\tablecaption{ \label{tab:conf_lim}Constraints on H$_{2}$O and OH Emission from Single-slab LTE Models.}
\tablewidth{500pt}
\tabletypesize{\small}
\tablehead{\colhead{Line Sample} & \colhead{Parameter} & \colhead{Quiescence (S05)} & \colhead{Outburst (A08)}}
\startdata
  \vspace{5pt}
& \textit{$T_{\rm ex}$} (K) & $580^{+40}_{-40}$ &  $595^{+130}_{-140}$ \\ %
 \vspace{5pt}
H$_{2}$O & \textit{$N_{\rm mol}$} ($10^{17}$ cm$^{-2}$) & $32.0^{+26.5}_{-13.7}$ &  $2.8^{+9.8}_{-1.7}$ \\ %
 & \textit{$r_{\rm p}$} (AU) & $0.58^{+0.07}_{-0.08}$ & (1.30)  \\ 

\\

\hline 
\\
  
  \vspace{5pt}
& \textit{$T_{\rm ex}$} (K) & (590) & $590^{+30}_{-20}$ \\
 \vspace{5pt}
low-$E_u$ OH & \textit{$N_{\rm mol}$} ($10^{17}$ cm$^{-2}$) & $\lesssim0.7$ & $2.8^{+0.6}_{-0.9}$ \\
 & \textit{$r_{\rm p}$} (AU) & (0.58) & $1.30^{+0.08}_{-0.08}$ 
\enddata
\tablecomments{The confidence limits on the parameters are derived using constant $\chi^{2}$ boundaries. In particular we report the $\Delta\chi^{2}=1.0$ boundaries, which in one-dimensional planes give the $68.3\%$ confidence on each parameter estimate. Values in parentheses are assumed from water and OH emission that are well characterized (see Section \ref{sec: res}). For OH we consider here only lines with $E_{u}\lesssim6000$ K (see Table \ref{tab:oh}).}
\end{deluxetable*}

\subsection{Summary of Changes in Emission from Model Results}
We report the constraints derived using single-slab LTE models in Table \ref{tab:conf_lim}. In Figures \ref{fig:models_qui} and \ref{fig:models_out} we show all together the LTE models derived for water and OH and the models assumed for organics. By comparison of the model results in quiescence and outburst we can summarize the changes in molecular emission as follows. Water becomes more optically thin and the column density decreases ($N_{\rm H_{2}O}$ from $\approx 3 \times 10^{18}$ to $\approx 3 \times 10^{17}$ cm$^{-2}$), but the emission comes from a larger area. Given the degeneracy between parameters, another possible solution is that the temperature increases while the column density remains the same over the same emitting area as in quiescence. However, the larger  emitting area of OH in outburst supports the former solution, assuming that the two molecules emit from the same portion of the disk \citep[as found e.g. by][]{dopp}. So, considering water and OH together we would conclude that the emitting area of warm gas increases from $\sim$ 0.6 to $\sim$ 1.3 AU in radius\footnote{Given the inclination of the disk of $\sim20^{\circ}$ \citep{sip} these projected equivalent radii may be tracing actual physical disk radii.}. Remarkably, while the column density of water - and of C$_{2}$H$_{2}$, HCN and CO$_{2}$ - decreases in outburst, the OH column density increases. OH emission can be explained by two temperature components, one characterized by the excitation of very high rotational levels ($E_u\gtrsim10,000$ K) and one by transitions with $E_u$ lower than $\sim$ 6000 K. The second component shows a temperature that is consistent with the result found for H$_{2}$O in outburst ($T_{\rm ex}\sim$ 600 K) and increases in column density in outburst for at least a factor 3 ($N_{\rm OH}$ from $\lesssim 0.7 \times 10^{17}$ to $\approx 3 \times 10^{17}$ cm$^{-2}$).

\section{IMPLICATIONS} \label{sec: disc}
Our study demonstrates that accretion outbursts in EX Lupi considerably affect the molecular chemistry in a region that is probably the surface of the disk at radii relevant for planet formation. If confirmed, a change in emitting area of OH and water would show that a larger extent of the disk is heated significantly during outburst. But a greater illumination and heating are not enough to explain the change in emission we observe, and we should consider the consequences of the spectral energy distribution of the emission. UV photodissociation of water is indeed the best candidate we find in the literature to explain the sudden formation of a large amount of OH in outburst. Moreover, if water becomes more optically thin in outburst and its column density decreases, this may allow UV radiation to penetrate deeper in the disk. Organics that might have been shielded by water in quiescence would not be so any more, and could be photodissociated. A thinner layer of water would also probably favor more emission from H$_{2}$ within disk regions that were previously not heated during quiescence. In the following we briefly discuss the scenario we propose to explain the change in molecular emission observed in EX Lupi between quiescence and outburst, and we compare our results with theoretical predictions from the literature.
\\

\subsection{H$_{2}$O and OH Emission between Quiescence and Outburst}
During the 2008 outburst in EX Lupi the mass accretion rate onto the star was found to increase from $\sim10^{-10}$--$10^{-9}$ to $\sim10^{-7}$ M$_{\odot}$ yr$^{-1}$ and the stellar+accretion luminosity $L_{*}+L_{acc}$ to increase by a factor $\sim$ 4.5 \citep{sip,asp,attila}. Whatever be the heating source of the gas in the disk, water formation via gas-phase reactions proceeds vigorously when the gas temperature is above 300 K \citep[e.g.][]{glas,bb09}, and can be expected to increase during an outburst. According to the \citet{glas} model, the range of H$_{2}$O column densities derived for EX Lupi in the two phases ($\approx 10^{17}$--$10^{18}$ cm$^{-2}$) can be explained with heating by X-rays irradiation alone, provided that H$_{2}$ formation on grains is efficient enough. However, it should be checked under which conditions formation of water would overcome its destruction under an increased X-ray radiation, and compare with the case of EX Lupi. Unfortunately, X-ray observations were not taken close enough in time to the \textit{Spitzer} spectra we consider in this work, and comparison of the available data from \citet{gud10} (August 2007) and \citet{grosso} (August 2008) does not provide conclusive constraints. \citet{glas} also proposed that accretion (mechanical) heating would increase the H$_{2}$O column density. This instead can be ruled out in EX Lupi: even if we had the same emitting area in quiescence and outburst the H$_{2}$O column density does not increase when the accretion rate is higher, more so if the emitting area is larger in outburst as we assume in our reference case (see Section \ref{sec: res1}).

On the other hand, UV radiation has been proposed to play a key role in the formation/destruction and availability of H$_{2}$O, OH, and organic molecules in the planet-forming zones of disks \citep{bb09}. A strong UV radiation can photodissociate water vapor in favor of OH in the upper layers of the warm disk atmosphere, competing with the capability of H$_{2}$O and OH to self-shield. The change in emission in EX Lupi is consistent with this scenario: we see $N_{\rm H_{2}O}$ decreasing in outburst, while $N_{\rm OH}$ increase up to the limit given by the self-shielding mechanism \citep[$\sim 2 \times 10^{17}$ cm$^{-2}$,][]{bb09}. Following in this direction, the increase of $N_{\rm OH}$ in outburst suggests that the connection between the strength of OH emission and the strength of the accretion rate found by \citet{cn11} may be attributed to the effect of UV radiation rather than to mechanical heating. Again, the available UV observations do not provide strong constraints but we have at least the evidence that UV radiation was higher at the time of the A08 spectrum than it was in quiescence, as in the U-band (0.36 $\mu$m) the observed flux varied by a factor $\sim50$ \citep{attila}. 

The strongest evidence in favor of a key role of UV radiation probably comes from OH emission. The coexistence of two temperature components in the MIR spectrum of EX Lupi is an interesting possibility. In previous works using \textit{Spitzer} spectra of classical T Tauri systems a cold component ($T \sim 500$ K) was found by \citet{cn08} fitting OH lines detected in the LH module, while \citet{cn11} proposed a hot component ($T \sim 4000$ K) considering lines in the SH module. Based on previous observational evidence, \citet{naji} proposed that UV-irradiated disks may show a twofold OH emission: a prompt non-thermal component from UV photodissociation of H$_{2}$O which would eventually relax to a thermal emission at the temperature of the gas. The broad population of OH levels in the range $E_{u} \approx$ 400-28,000 K, first observed by \citet{tap} in the HH211 outflow, was indeed explained with a radiative cascade down the ground-state rotational ladders following UV photodissociation of H$_{2}$O, which would preferentially populate OH levels with E$_{u} \gtrsim 40,000$ K \citep{tap,har}. If our approximation of two temperature components in outburst is confirmed, it would represent the first observational evidence of an ongoing prompt ``hot" (in reality non-thermal) emission followed by a thermal emission at the temperature of the gas. In fact, in outburst we find one component that may hardly be explained with thermal excitation, while another component shows the same temperature found for water (see Section \ref{sec: res2}). An interesting peculiarity of EX Lupi is the detection of the cross-ladder transitions between $\sim$ 15 and 19 $\mu$m, that were never observed before in any classical T Tauri system.  The case of EX Lupi, as well as similar variable sources, might be very useful to clarify the nature of OH emission in the inner regions of circumstellar disks, but more theoretical work is needed. For instance, it would be interesting to know if the proposed excitation mechanism by radiative cascade would produce the curve that we see in the rotation diagram, and, if that is a transitory process, on which timescales/conditions it should be observed as two components rather than a continuum of temperatures.

The simultaneous monitoring of UV, X-rays and MIR emission during future outbursts would help in confirming the connection between the variation in accretion rate and changes in H$_{2}$O and OH emission. Velocity- and transition-resolved spectral lines might be the key tool to clarify the location of the MIR molecular emission as well as help in understanding the processes responsible for formation/destruction of the molecules we observe \citep[e.g.][]{pont10b}. More comprehensive chemical disk models accounting for both X-rays and UV variable irradiation would also be very helpful in better constraining water formation and survival in the inner zones of disks based on spectrally resolved synoptic observations of young variable star+disk systems.

\subsection{The Lack of C$_{2}$H$_{2}$, HCN and CO$_{2}$ in Outburst}
The fact that organic emission dicreases in outburst is very interesting, and probably the most puzzling of our findings. Production of HCN has been proposed to be driven by strong UV photodissociating N$_2$ in disk atmospheres \citep{pasc09}. In EX Lupi we observe instead a lower HCN column density in outburst, and yet the ratio $N_{\rm HCN} / N_{\rm C_{2}H_{2}} \sim 2.7$ in quiescence is consistent with the median sun-like star reported in \citet{pasc09}. The lack of C$_{2}$H$_{2}$, HCN and CO$_{2}$ in outburst seems to preferentially imply the large photodissociation of these molecules by UV, in contrast with the capability of self-shielding of H$_{2}$O and OH. If the organic emission probes the same disk radii as H$_{2}$O and OH, then their non-detection would show that warm organics are located higher in the disk atmosphere than an optically thick layer that could shield them, whether it is made of dust or H$_{2}$O and OH as suggested in \citet{pont10} and \citet{bb09} respectively. The unclear behavior of the emission from these organic molecules in other T Tauri disks \citep{cn11} suggests that we still do not fully understand where exactly their emission originates and what are the factors most responsible for their abundance.
It would be very interesting to see if organics are detected in EX Lupi again after the outburst. If that was found to be true, than the vertical mixing proposed by \cite{attila} to be powered-up in outburst may be the mechanism replenishing the disk atmosphere with molecules from closer to the midplane. Monitoring EX Lupi during the current new quiescent phase and future outbursts may be extremely important to check the timescales/processes relevant for organic chemistry in planet-forming regions of disks. 

\subsection{Note on HI and H$_{2}$ in Outburst}
We do not focus in this paper on the strong lines from H \textsc{i} and H$_{2}$ that appear in EX Lupi in outburst (see Section \ref{sec: anal1}). The increase in the accretion rate is certainly a possible explanation for an increase in H \textsc{i} emission \citep[e.g.][]{pasc07}, but photoevaporative winds might also contribute \citep[see e.g. the discussion in][]{naji}. The increase in H$_{2}$ emission can be explained by emission from a warm disk atmosphere \citep{naji,gorti}, which in EX Lupi in outburst is probably warmed up in regions that were too cold in quiescence (larger radii and/or deeper layers). It would be interesting to see whether a direct relation between these species and the formation/destruction of H$_{2}$O in EX Lupi exists. If warm H$_{2}$ is indeed available on a larger portion of the disk atmosphere, then water vapor formation might be favored \citep{glas} and increase again the observed column density after the end of the outburst, when UV photodissociation diminishes.

\section{CONCLUDING REMARKS}
The molecular emission we detect in the quiescent S05 spectrum (from H$_{2}$O, OH, HCN, C$_{2}$H$_{2}$ and CO$_{2}$) is comparable to what has been found to be common in several other T Tauri systems \citep{pasc09,cn08,cn11}. This is a further confirmation that EX Lupi is not distinguishable from typical T Tauri systems, as already suggested by \citet{sip} from consideration of the spectral energy distribution in quiescence. Therefore, the remarkable changes in molecular emission observed in the outbursting EX Lupi might be extremely important to constrain chemical-physical processes common to all T Tauri systems. In this paper we have addressed the simple scenario of the increase of one system parameter in EX Lupi: the disk-atmosphere illumination during outburst. To first order this is consistent with an outbursting event that accreted material only from inside an inner hole in the dusty disk ($\sim0.3$ AU) and not over global disk scales \citep{attila,goto,kosp11}. Consideration of more sophisticated scenarios needs to be included as soon as high-resolution data will become available. The geometrical structure of the EX Lupi system is probably affected by strong outbursts and may be relevant for the correct interpretation of the changes in emission we observe, including the increase of line fluxes in the LH module. As the stellar+accretion luminosity $L_{*}+L_{acc}$ increased in outburst by a factor $\sim$ 4.5 \citep{asp}, the dust-sublimation radius is probably shifted outward by a factor $\sim$ 2. This can have effects on the inner rim location and the shadowed portion of the disk as well, and/or the location of the snow lines of the different chemical species. The location of the disk we likely probe, the inner few AUs, is indeed where both the shadowing and the snow line are believed to be. In addition to that, UV photodesorption from icy grains is expected to power the production of OH as well as to largely replenish water vapor in the disk atmosphere \citep{oberg}. Despite the increased UV and the production of OH, we still see a large column density of water vapor in EX Lupi in outburst. If self-shielding of water breaks under particularly harsh conditions (see the case of Herbig AeBe stars, e.g. \citet{pont10} and \citet{fed}, and transitional disks, e.g. \citet{naji}), then a receding snow line may still provide the ice reservoir needed to sustain a wet disk atmosphere in EX Lupi. Velocity-resolved observations are critical to confirm the location of the different molecular emissions and explore these senarios. New infrared facilities from the ground (e.g. the soon upgraded VISIR on the \textit{Very Large Telescope}, and in the future METIS on the \textit{E-ELT}) and from space (with MIRI on the \textit{James Webb Space Telescope} and MIRES on \textit{SPICA}) will be the key tool in future investigations of the warm molecular emission from the inner regions of disks.
\\

We are very pleased to acknowledge the anonymous referee for comments that helped in improving considerably this work, as well as the several colleagues who contributed to our investigation with valuable discussions, in particular Ted Bergin, Klaus Pontoppidan, Inga Kamp, John Carr, and Joan Najita. AB wants to thank all people of the new Star and Planet Formation group at the ETH Zurich for their support during the development of this work, especially Susanne Wampfler and Michiel Cottaar. This work is based on observations made with the \textit{Spitzer Space Telescope}, which is operated by the Jet Propulsion Laboratory, California Institute of Technology.

\appendix
\section{A Single-Slab LTE Model} \label{app: model}
A single-slab model is used to extract from the spectra the excitation conditions of the emitting gas. We assume the gas to be in local thermodynamical equilibrium (LTE), but account for optical depth effects. We follow a formalism similar to \citet{vdt07}, and we ignore the absorption of background radiation by the emitting gas (i.e. the absorption term of the radiative transfer equation). Neglecting mutual overlap between transitions and assuming the line profile function to be Gaussian, the integrated intensity (e.g. erg s$^{-1}$ cm$^{-2}$ sr$^{-1}$) of a transition connecting the molecular level $u$ (upper) and $l$ (lower) reads
\begin{equation} \label{eq:intens}
I_{ul} = \frac{1}{2\sqrt{\ln(2)}} \, \frac{\Delta v \nu_{ul}}{c}  \,  B_\nu(T_{\rm ex}) \int_{-\infty}^{\infty} 1 - \exp\left(-\tau_{ul} e^{-y^2}\right)\, dy  \, .
\end{equation}
Here, $\tau_{ul}$ is the opacity at the line center, $B_\nu(T_{\rm ex})$ the Planck function for the excitation temperature $T_{\rm ex}$, $\nu_{ul}$ the line frequency  and $\Delta v$ the line width, which is fixed to 1 km s$^{-1}$. The integral in the above expression has been obtained using a change in variable $y=2 \sqrt{\ln(2)} v/\Delta v$ and approaches $\sqrt{\pi}\left(1-e^{-\tau_{ul}}\right)$ for $\tau_{ul} < 1$. Thus, the usual expression used in escape probability codes \citep[e.g.][]{vdt07} is recovered for low opacities. The opacity at the line center is given by 
\begin{equation} \label{eq:tau}
\tau_{ul} = \frac{\sqrt{\ln(2)}}{4 \pi \sqrt{\pi}}  \frac{A_{ul} N_{\rm mol} c^3}{ \Delta v \nu_{ul}^3} \, \left(x_l \frac{g_u}{g_l} - x_u \right) \, ,
\end{equation}
with the Einstein-A coefficient of the transition $A_{ul}$ and the total molecular column density $N_{\rm mol}$. The statistical weight and the normalized level population ($\sum_i x_i = 1$) of the upper and lower level are given by $g_u$, $g_l$ and $x_u$, $x_l$, respectively. The level population is obtained from the Boltzmann distribution $x_i = g_i \exp\left(-E_i/k T_{\rm ex} \right) / Q(T_{\rm ex})$, with the partition sum $Q(T_{\rm ex})$ and the energy of the level $E_i$. The relevant molecular parameters are taken from the HITRAN 2008 database \citep{hitran}.

To model the Spitzer spectra, {\it i.)} the intensity of all transitions of a chosen molecule (e.g. H$_{2}$O) in the wavelength range of the \textit{Spitzer} IRS is calculated using equations \ref{eq:intens} and \ref{eq:tau}, for an assumed $T_{\rm ex}$ and $N_{\rm mol}$, {\it ii.)} the integrated flux ($F_{ul}$, in units of erg s$^{-1}$ cm$^{-2}$) is calculated by multiplying the intensity $I_{ul}$ with the solid angle $\Delta \Omega= A / d^2$, where $A$ is the emitting area in the sky, for which we assume a projected circle with radius $r_{\rm p}=\sqrt{A/ \pi } $, and $d$ the distance to the source \citep[assumed to be consistent with the distance of the Lupus complex, 150 pc,][]{lomb}; {\it iii.)} a synthetic spectrum (in erg s$^{-1}$ cm$^{-2}$ $\mu$m$^{-1}$) with spectral resolution $R=\Delta\lambda/\lambda \approx 600$ is generated based on all transitions in the wavelength range considered, with flux given by
\begin{equation}
F(\lambda) = \sum_i  \frac{1}{\sqrt{2\pi} \sigma_\lambda} e^{-\left(\frac{\lambda-\lambda^i_{ul}}{\sqrt{2}\sigma_\lambda}\right)^2} F^i_{ul} \ ,
\end{equation}
where $\sigma_\lambda = \lambda / 2 R \sqrt{2 \ln(2)}$, and $\lambda^i_{ul}$ and $F^i_{ul}$ are respectively the wavelength and the flux of a single transition.

\end{document}